\documentclass[sigplan, screen, authorversion, 10pt]{acmart}

\AtBeginDocument{%
  \providecommand\BibTeX{{%
    \normalfont B\kern-0.5em{\scshape i\kern-0.25em b}\kern-0.8em\TeX}}}
    
\usepackage[os=win]{menukeys}
\usepackage{listings}
\usepackage{array}
\usepackage{microtype}

\newcommand{\mytilde}{\raise.17ex\hbox{$\scriptstyle\sim$}}

\definecolor{backgroundGrey}{HTML}{F2F2F2}
\newcommand{\changed}[1]{#1}

\newcolumntype{L}[1]{>{\raggedright\let\newline\\\arraybackslash\hspace{0pt}}m{#1}}
\newcolumntype{C}[1]{>{\centering\let\newline\\\arraybackslash\hspace{0pt}}m{#1}}
\newcolumntype{R}[1]{>{\raggedleft\let\newline\\\arraybackslash\hspace{0pt}}m{#1}}

\setcopyright{acmlicensed}
\acmPrice{15.00}
\acmDOI{10.1145/3563835.3567663}
\acmYear{2022}
\copyrightyear{2022}
\acmSubmissionID{onward22papers-p78-p}
\acmISBN{978-1-4503-9909-8/22/12}
\acmConference[Onward! '22]{Proceedings of the 2022 ACM SIGPLAN International Symposium on New Ideas, New Paradigms, and Reflections on Programming and Software}{December 8--10, 2022}{Auckland, New Zealand}
\acmBooktitle{Proceedings of the 2022 ACM SIGPLAN International Symposium on New Ideas, New Paradigms, and Reflections on Programming and Software (Onward! '22), December 8--10, 2022, Auckland, New Zealand}

\begin{document}

\title{Forest: Structural Code Editing with Multiple Cursors}

\author{Philippe Voinov}
\email{philippevoinov@gmail.com}
\affiliation{%
    \institution{ETH Zurich}
    \country{Switzerland}
}
\author{Manuel Rigger}
\email{rigger@nus.edu.sg}
\affiliation{%
    \institution{National University of Singapore}
    \country{Singapore}
}
\author{Zhendong Su}
\email{zhendong.su@inf.ethz.ch}
\affiliation{%
    \institution{ETH Zurich}
    \country{Switzerland}
}


\begin{abstract}
Software developers frequently refactor code.
Often, a single logical refactoring change involves changing multiple related components in a source base such as renaming each occurrence of a variable or function.
While many code editors can perform such common and generic refactorings, they do not support more complex refactorings or those that are specific to a given code base.
For those, as a flexible---albeit less interactive---alternative, developers can write refactoring scripts that can implement arbitrarily complex logic by manipulating the program's tree representation.
In this work, we present Forest, a structural code editor that aims to bridge the gap between the interactiveness of code editors and the expressiveness of refactoring scripts.
While structural editors have occupied a niche as general code editors, the key insight of this work is that they enable a novel structural multi-cursor design that allows Forest to reach a similar expressiveness as refactoring scripts; Forest allows to perform a single action simultaneously in multiple program locations and thus support complex refactorings.
To support interactivity, Forest provides features typical for text code editors such as writing and displaying the program through its textual representation.
Our evaluation demonstrates that Forest allows performing edits similar to those from refactoring scripts, while still being interactive.
We attempted to perform edits from 48 real-world refactoring scripts using Forest and found that 11 were possible, while another 17 would be possible with added features.
We believe that a multi-cursor setting plays to the strengths of structural editing, since it benefits from reliable and expressive commands. Our results suggest that multi-cursor structural editors could be practical for performing small-scale specialized refactorings.
\end{abstract}

\ccsdesc[500]{Software and its engineering~Integrated and visual development environments}
\ccsdesc[300]{Software and its engineering~Software evolution}
\ccsdesc[300]{Software and its engineering~Software maintenance tools}

\keywords{Structural editing, refactoring, multi-cursor}

\maketitle

\section{Introduction}
\label{sec:introduction}
When maintaining and extending software, developers are often forced to make repetitive edits. For example, when a developer introduces a new required parameter to a function, a corresponding argument has to be added at every call site. Similarly, changes such as splitting a large module or class into multiple parts require repetitive adjustments at most usage sites. If the changes affect systems that are used in many locations in a codebase, such as logging or database access, refactoring can be especially time-consuming.

\begin{figure}[!b]
    \includegraphics[width=\columnwidth]{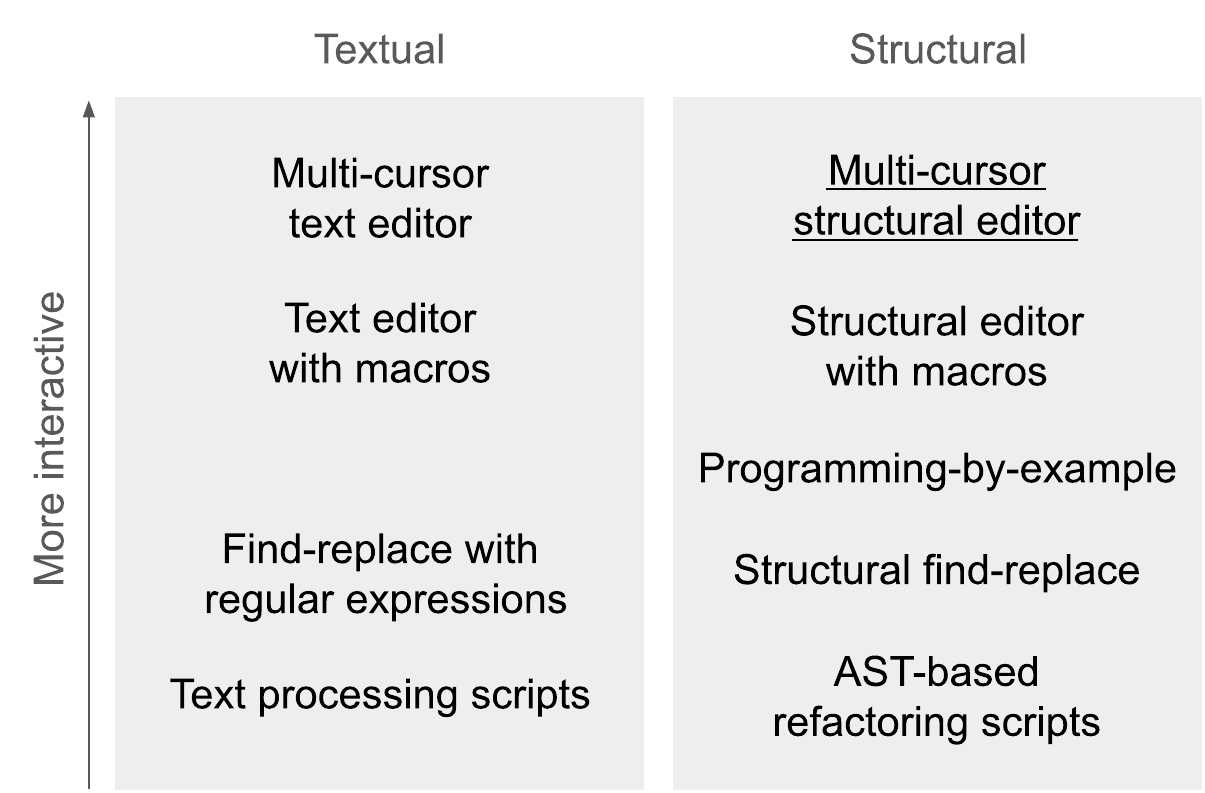}
    \captionof{figure}{The design space of tools for performing similar code edits in multiple locations. Forest is the first tool in the category of multi-cursor structural editors (underlined).}
    \label{fig:interactive-textual-structural-space}
\end{figure}

Various approaches have been proposed to support repetitive refactoring on both textual and structural level (see Figure \ref{fig:interactive-textual-structural-space}). These tools differ in terms of interactivity and whether they can be classified as textual or structural. We consider a tool more interactive when it provides a workflow similar to conventional single location code editing and immediately shows results in all locations. While textual tools work with characters in a text file, without an understanding of the programming language in question, structural tools parse the program and operate on its Abstract Syntax Tree (AST). For some common edits, such as renaming a variable and all usages, Integrated Development Environments (IDEs) contain built-in refactorings that operate structurally. Existing tools are discussed in detail in Section \ref{sec:related-work}.


Textual tools (left half of Figure \ref{fig:interactive-textual-structural-space}), such as multi-cursor text editors or regular expressions, have many advantages over structural tools. They use familiar concepts from single-location text editing. Since these tools do not inspect the syntax of a program, they work with any programming language and in programs with invalid syntax. However, this lack of syntactic understanding makes it complex or impossible to perform certain edits in multiple locations. For example, since many operations in textual tools depend on the way that a program is formatted, they require the user to make adjustments to match their code style. Attempting to use find-replace with regular expressions to swap the order of two arguments in every call to a function highlights these limitations: If the function calls are always on a single line and the arguments themselves are syntactically simple (e.g. numeric literals), then the regular expression is simple. However, in the general case where a function call spans multiple lines and its arguments may themselves be function calls, correctly performing the edit would require parsing the code. Multi-cursor text editors, editors where every movement and text edit is replayed in multiple locations simultaneously, provide an interactive editing experience for multiple locations. They allow users to edit multiple locations using the same commands as for a single location, letting users apply familiar concepts from conventional text editors. However, the edits that they can express are limited. Typically, only commands such as inserting text or moving in whole words or lines are supported.

Structural tools (right half of Figure \ref{fig:interactive-textual-structural-space}) can perform more complex edits than textual tools, and provide commands which can be reliably applied in many locations. However, they only work with valid syntax and require explicit support for each programming language. AST-based refactoring scripts are programs in a general purpose programming language that directly manipulate an AST. They are used to perform repetitive edits in large scale code bases. At this scale, operating on an AST is significantly more reliably than textual operations. Additionally, since such scripts use a general purpose programming language, they can encode more complex editing logic than other tools. However, the process of developing such scripts is completely different to conventional code editing (the user must write a program to edit their code, instead of editing it directly), making them impractical for small-scale repetitive edits. Structural find-replace tools are the structural equivalent of find-replace with regular expressions. They are more reliable than textual find-replace, but require an understanding of the AST and use a separate set of concepts than conventional text editing. Additionally, both textual and structural find-replace have limited support for encoding logic, such as filtering which locations should be edited.


At a high level, we propose to combine multi-cursor editing with structural editing commands. This preserves the key advantages of multi-cursor text editors: multiple locations are edited in the same way as a single location and results are immediately shown. However, by using structural commands instead of textual commands, more complex edits are supported, and edits can be performed reliably in many locations regardless of formatting and nested syntax.

Our prototype editor, \emph{Forest}, is a multi-cursor structural editor for TypeScript. In Forest, cursors point to AST nodes. The user navigates within the AST by using structural operations like ``move to parent''. These operations work reliably in situations where textual tools are ineffective, such as with complex nested syntax. To retain the strengths and familiarity of text editors, insertions are performed by parsing typed text. A cursor can be split (e.g., for each child of the selected AST node) to create multiple cursors that handle editing commands simultaneously. This allows users to perform repetitive edits the same way as they would when editing a single location. A novel \emph{hierarchy of cursors} concept makes it practical to work with cursors that were split multiple times.

\begin{figure*}[t]
    \includegraphics[width=\linewidth]{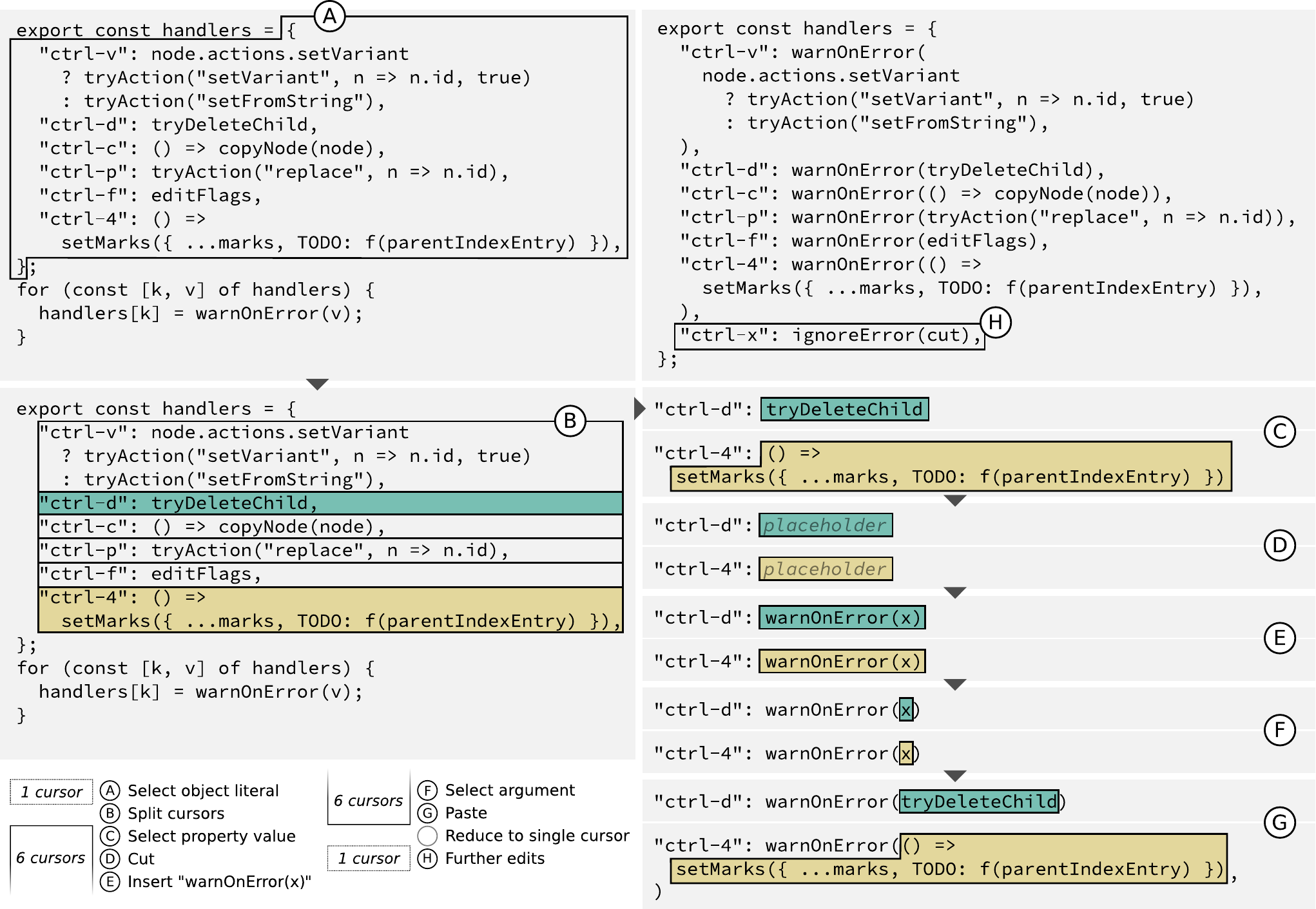}
    \captionof{figure}{A repetitive code edit where the user starts with the code in the top left and wishes to arrive at the code in the top right. Each circled letter describes a step of this edit as performed in Forest. Each grey box of code marked by a letter shows the state after the described edit was performed. The black outlines are cursors (multiple cursors can exist simultaneously in one step). Steps C through G only show two of the cursors for brevity.}
    \label{fig:motivational-example}
\end{figure*}

To understand the strengths and weaknesses of multi-cursor structural editing, we collected real-world AST-based refactoring scripts and attempted to perform the corresponding edits interactively in Forest. Of the 48 edits, 11 could be performed without any significant issues, and a further 17 would likely become practical with improvements to the editor. This shows that developers could avoid writing some---but not all---refactoring scripts and instead use a multi-cursor structural editor to perform their edit interactively. 

We make three main contributions:
\begin{itemize}
    \item We built Forest\changed{\footnote{\label{footnote:forest-demo-site}\changed{Online version of Forest with interactive examples:\newline \url{https://forestonward2022.walr.is/}\newline{}GitHub: \url{https://github.com/tehwalris/forest}\newline{}Archive: \url{https://doi.org/10.5281/zenodo.7225442}}}} --- a structural editor prototype for TypeScript. Forest is one of very few structural editors for modern languages with complex syntax.
    \item We extended Forest to support multiple cursors, with special integration between multi-cursor editing and structural editing. This is the only multi-cursor structural editor with such integration.
    \item We evaluated how Forest compares to refactoring\linebreak{}scripts, showing that Forest can be used to perform edits from some real-world refactoring scripts.
\end{itemize}
\pagebreak

\section{Motivating Example}
\label{sec:motivating-example}

Figure \ref{fig:motivational-example} shows a repetitive code edit which we will use to motivate and demonstrate multi-cursor structural editing. This is an edit to TypeScript code in an old version of Forest's own codebase. We wanted to wrap every property value (the expressions to the right of \lstinline$:$ within the object literal \lstinline${ ... }$) in a function call (\lstinline$warnOnError$) directly, rather than using a loop to wrap the values at run time. We made this change to subsequently add a new property which, unlike the others, had to be wrapped in \lstinline$ignoreError$.

Textual approaches cannot reliably perform such a refactoring, because the task is inherently not textual. Selecting every property is especially difficult, because they span varying numbers of lines and the property separator comma appears deep inside the values of some properties. In order to perform the task using textual find-replace, a regular expression that captures the property names and values is required. A reasonable attempt might be:
\begin{lstlisting}
(".*?"): (.*?),\n(?=  "|})
\end{lstlisting}
However, such a regular expression relies on the exact number of spaces used for indention in order to discern which commas separate properties and which are part of deeply nested expressions. A different or inconsistently applied code style would break this approach. The same issue occurs in multi-cursor text editors when the user creates multiple cursors by splitting an existing cursor (marked A in Figure \ref{fig:motivational-example}) using a regular expression that describes the separator (\lstinline$,(?=\n  ")$).

Structural tools can perform this refactoring reliably, avoiding text-related issues like identifying separators. However, there are reasons that discourage the use of existing structural tools for such a task. The developer could write a refactoring script ~\cite{noauthor_jscodeshift_2021} that transforms the Abstract Syntax Tree (AST) of the program. However, writing a refactoring script is unreasonably heavyweight for an edit that is required in so few locations. This edit is also too program-specific to be available as a built-in refactor in an Integrated Development Environment (IDE). Programming-by-demonstration tools ~\cite{kim_recommending_2014}~\cite{ni_recode_2021}~\cite{kim_recommending_2014}, which infer a structural edit based on an example in one location, are a more lightweight solution than writing a custom refactoring script. However, such tools may not always perform the desired edit, since the inference process may either fail or may misinterpret the given examples.


Forest allows the user to perform the example refactoring with the same reliability as other structural tools, while retaining the interactivity of multi-cursor text editing. Using Forest in Figure \ref{fig:motivational-example}, the user starts their edit by moving their cursor to the object literal containing all the properties that they want to modify. This cursor is shown as a black outline in step A of Figure \ref{fig:motivational-example}. This cursor points to the AST node that represents the object literal, but is shown to the user as a selected text range (see Section \ref{sec:selections} for details).

The user then applies Forest's "split cursors" command, which replaces the single cursor pointing to the object literal by multiple cursors --- one cursor for every node in the AST that is a child of the object literal. The new cursors are shown as black outlines in step B of Figure \ref{fig:motivational-example}. The operation of finding the boundaries between properties is the key operation in this example refactoring that structural tools can easily perform, but that textual tools will struggle with. 

In steps C through G, multiple cursors are used. For brevity, we only show two of the six cursors. Each command issued by the user is processed by all cursors simultaneously. This allows the user to perform their edit exactly the same way as if they were editing a single location. This is a major advantage of multi-cursor editing in general (not just of multi-cursor \emph{structural} editing) over tools like refactoring scripts, regular expressions, and structural find-replace, which are very different from tools for editing single locations. 

In step C, the user moves the selection (of each cursor) from the object property to its last child in the AST. This is a structural navigation operation, which would work reliably even if the left hand side of the property had more complicated content (e.g., a computed property name in TypeScript). Steps C through G use structural navigation, placeholders, copy-paste, and continuous pretty printing. These are common operations in Forest. They are described in Section \ref{sec:single-cursor-overview}.

This example mostly benefited from the structural nature of Forest to split the cursors. Most of the edits in steps C through G could be performed as easily in a multi-cursor \emph{text} editor as in Forest. In Section \ref{sec:multi-cursor} we discuss operations in Forest that are unique to multi-cursor \emph{structural} editors and that are useful once the cursor has already been split.

\section{Forest}
\label{sec:single-cursor}
This section describes Forest, but without any of its features related to multi-cursor editing. Our main conceptual contributions are concerned with multi-cursor editing, which is discussed separately in Section \ref{sec:multi-cursor}. The section establishes the necessary context to understand our core contribution. \changed{We have recorded a short supplementary video (part of our archived artifacts --- see Footnote \ref{footnote:forest-demo-site}) to give an intuition of how Forest's basic features can be used.}

\subsection{Overview of Design}
\label{sec:single-cursor-overview}
Source files in Forest are shown as pretty-printed text. This text is synchronized with a Forest-specific AST (referred to as ``tree'' from now on) which the user can interact with. Forest is a modal editor --- it has a \emph{normal} mode for navigating and issuing commands and an \emph{insert} mode for inserting text. A full list of available commands is given in the appendix.

\paragraph{Navigation} The user controls a cursor that always has part of the source file selected. This selection must correspond to a contiguous selection of siblings in the tree. Forest provides structural navigation commands (e.g., ``go to parent'' or ``select next leaf node'') for changing the selection. The navigation commands treat every node in the tree as a list of children in text order. This makes it possible to navigate through any node using the same commands.

\paragraph{Insertion} To add to the source file, the user enters \emph{insert mode} at the text location at the start (or end) of their selection, types normal source code, and exits insert mode. It is only possible to exit insert mode when the source file has valid syntax. This restriction simplifies implementation and is common in structural editors. 


\paragraph{Modification} Existing code can be modified using the delete, copy and paste commands. These commands are structural: they modify the tree (not the text) directly, after which the modified tree is pretty-printed and shown to the user. Deleting a node may result in a tree that does not correspond to a valid TypeScript AST (e.g., deleting \lstinline{b} in \lstinline{a + b}). In this case, Forest adds a placeholder (effectively a hole) instead of the deleted item.



\paragraph{AST} Most nodes in the Forest tree correspond exactly to the equivalent TypeScript AST node. However, for nodes where the TypeScript AST is inconvenient to edit, Forest uses a different structure. For example, chains of property accesses and calls (\lstinline{this.data.filter(...).map(...)}) are a left-associative tree in TypeScript, but a single flat list in Forest (\lstinline{this}, \lstinline{data}, \lstinline{filter}, \lstinline{(...)}, \lstinline{map}, \lstinline{(...)}).

\subsection{Selections}
\label{sec:selections}
The behavior of selections in Forest has been significantly influenced by the ABC editor \cite{meertens_abc_1992}. A selection (the \emph{focus}) is one or more nodes in the tree, which must be contiguous siblings. Any selections with the same text range are considered equivalent. \changed{Note that although the selection of a single cursor must consist of continuous siblings, there is no restriction that multiple cursors must have selections that are adjacent to each other.}

\sloppy{}
\paragraph{Equivalent Selections} We will use the \lstinline{TypeReferenceNode} from the TypeScript AST to demonstrate equivalent selections. This node is used to refer to a previously declared type. It contains an \lstinline{Identifier} (the referenced type) and an optional list of type parameters. Depending on whether the type parameter list is used, a \lstinline{TypeReferenceNode} looks like \lstinline{MyGenericType<T>} or \lstinline{MyNonGenericType}. Note that when the type parameter list is not used, selecting the \lstinline{TypeReferenceNode} would have the same text range as selecting its inner \lstinline{Identifier}. Forest treats selections with equal text range as one, meaning it is ambiguous which AST node is selected. However, any command that is applied will effectively disambiguate the selection. In our example, if ``move to parent'' is performed, then the selection will move to the parent of the \lstinline{TypeReferenceNode}, but if ``rename'' is performed, then the command will affect the \lstinline{Identifier}. By not allowing the user to make two different selections with the same text range, we hope to minimize confusion about what is selected. This also eases navigation, since the tree (as perceived by the user) is less deep.

\paragraph{Selections and Text Insertion} The design of equivalent selections in Forest is closely tied to text insertion, since text insertion effectively disambiguates equivalent selections. Consider the function call \lstinline{f(x.y)} with the argument \lstinline{x.y} focused. It is ambiguous whether the argument \lstinline{x.y} itself (containing \lstinline{x} and \lstinline{y}) or the list of arguments to \lstinline{f} (containing the single argument \lstinline{x.y}) is focused. If we had used a non-text-based command for inserting list items (like ``append item to list'' followed by a menu selection), we would need a way to disambiguate which list the user wishes to append to. However, a text insertion naturally disambiguates this: If the user chooses to append \lstinline{.z} (beginning with a period), it is clear that they are appending a property access to the argument itself. If they choose to append \lstinline{,z} (beginning with a comma), it is clear that they are appending a new argument to the argument list of \lstinline{f}.

\subsection{Flatter AST}
\label{sec:ast-adjustments}
Most types of syntax are represented in the Forest tree the same way as in the TypeScript AST. However, some structures which are represented by binary trees in the TypeScript AST are represented as flat lists in Forest. We believe that it is difficult for users to imagine the underlying binary tree when looking at textual code. 

\paragraph{Loose and Tight Expressions} We convert typical expressions to two levels of flat lists, which we call \emph{loose expressions} (whose elements are generally separated by spaces) and \emph{tight expressions} (generally printed without whitespace). For example, \lstinline{a.x() + f.g * c} is a loose expression containing \lstinline{a.x()}, \lstinline{+}, \lstinline{f.g}, \lstinline{*}, and \lstinline{c}. Similarly, \lstinline{a.x()} is a tight expression containing \lstinline{a}, \lstinline{x}, and \lstinline{()}. 
The separation of loose and tight expressions seems intuitive and predictable to us, but we have not investigated this design choice with users.

\paragraph{Operator Precedence} Forest's adjusted tree with flattened lists loses precedence information. This is done intentionally to simplify navigation. It also allows replacing operators in a way that ignores precedence, like in a text editor. The ABC editor also flattens expressions around binary operators and ignores precedence.

\paragraph{Typical TypeScript Edits} Having expressions flattened at the level of tight expressions is helpful for edits that commonly appear when writing TypeScript. For example, consider the following expression:
\begin{lstlisting}
myArray.map(x => x + 1).filter(x => x < 0)
\end{lstlisting}
Since Forest interprets this expression with a flat structure (\lstinline{myArray}, \lstinline{map}, \lstinline{(x => ...)}, \lstinline{filter}, \lstinline{(x => ...)}) the user can easily perform the following operations:
\begin{itemize}
    \item Navigate through the chained calls (without imagining the binary tree),
    \item remove calls (even from the middle of the chain, e.g., \lstinline{map(...)}),
    \item add new identifiers to form property accesses (e.g., prepend \lstinline{this.} to the whole chain),
    \item add or remove the function calls themselves (e.g., add \lstinline{join} to the end of the expression, resulting in an unevaluated function, then later add the missing \lstinline{()}, resulting in a string),
    \item add inline operators like \lstinline{?.} (optional chaining) (e.g., add \lstinline{?.} after \lstinline{myArray} in case it is \lstinline{undefined}).
\end{itemize}

\subsection{Navigation}
Forest has a small set of navigation commands, most of which are directional in a tree sense (e.g., going \emph{up} towards the parent, or going to the \emph{next} sibling). None of the navigation commands depend on the way the code is formatted.

\paragraph{Dependency on Formatting} 
In text editors (e.g. Vim), the user must issue different editing commands depending on how the code was pretty-printed. For example, to delete a function argument one would use ``delete up to including comma'' (not last argument), ``delete to closing parenthesis'' (last argument), or ``delete line'' (formatting with one argument per line). Forest makes navigation be completely independent of pretty-printing, which is only possible in a structural editor. This requires limitations (e.g., moving down by a few lines is not possible in general), but it also allows for special integration with the pretty-printer.

\paragraph{Continuous Pretty-Printing} The textual code displayed in Forest is pretty-printed after every single edit. This frees the user from thinking about minor details of code formatting, since it is not possible---not even temporarily---to change the style of their code. Since Forest's navigation does not depend on how code is pretty-printed, it is possible to run the pretty-printer \emph{asynchronously}. For example, the user could perform an insertion and then start navigating, while the editor performs a pretty-print in the background (without blocking the user's navigation), and eventually shows the new print. This would not be possible if Forest had commands like ``delete line'', because if the user issues this command right before the editor switches to a new print, the meaning of the command would change and might not do what the user intended.


\paragraph{Empty Selections}
Forest does not allow empty selections, except in the following case: Consider the function call \lstinline{f(x, y)} and a selection inside the argument list (covering \lstinline{x, y}). The user can append a function argument using the ``insert text after cursor'' command. However, if the function call had no arguments, then the user could not focus the argument list, since it would be an empty selection. To make the situation the same with no arguments as with some arguments, we allow selecting the content of the empty parenthesized lists.


\subsection{Deletion and Placeholders}
\label{sec:deletion}
As the tree that the user edits in Forest is slightly different from the TypeScript AST, deleting a node may make Forest's tree no longer correspond to a valid TypeScript AST. For example, \lstinline{f()} is a tight expression consisting of \lstinline{f} and \lstinline{()} in Forest. Deleting \lstinline{f} is reasonable (e.g., to replace it with another identifier), but the remainder \lstinline{()} does not correspond to a valid TypeScript AST for an expression. In this case, Forest inserts the identifier \lstinline{placeholder} in place of the deleted identifier. This allows all existing tooling (the TypeScript parser and any pretty-printers) to work normally. Forest tracks the fact that this node is a placeholder across edits and pretty-printing, so that special highlighting and behavior can be provided for placeholders. Forest's placeholders are effectively holes. 


\section{Multi-Cursor Editing}
\label{sec:multi-cursor}
Forest is the first structural editor that is specifically designed for multi-cursor editing. We introduce multiple novel features including a hierarchy of cursors, switchable multi-cursor modes, and marks that track AST nodes. In this section, we describe the features of Forest which are either specific to multi-cursor editing or are especially useful with multiple cursors. A full list of multi-cursor editing commands is given in the appendix.


\subsection{Relaxed Mode}
When multiple cursors exist in Forest, each movement or editing command performed by the user is executed by every cursor simultaneously. The clipboard and some other editor state is stored per cursor. If some cursor can not handle a command (e.g., move one character to the right, but the cursor is already at the end of the document), no special handling is performed. The single failing cursor will simply do nothing, while the other cursors will still process the command. This ``broadcast command and ignore failures'' approach is called \emph{relaxed mode} in Forest. Unlike other multi-cursor editors, Forest has alternative \emph{multi-cursor modes}, which give different behavior (Section \ref{sec:filtering-cursors}).

\subsection{Creating Multiple Cursors}
\paragraph{Manually} The most direct way to obtain multiple cursors is to manually mark where they should be created. In Forest, the user moves the cursor to the desired selection and issues a command to \emph{queue} it. After all desired locations have been queued, the user can switch to the queued set of cursors.

\paragraph{Splitting} In Forest, a selection containing multiple nodes can be split using a single command. Forest will create a new cursor for every child of the AST node that is currently selected. In contrast to other editors, the user does not need to supply any further information (e.g., regular expressions describing separators in Kakoune \cite{coste_kakoune_nodate}). The AST naturally defines how the cursor should be split. This is especially useful for splitting complex nested expressions, since separators in the inner expressions are perfectly ignored by the structural split operation, but would be difficult to ignore using regular expressions or similar approaches.

\paragraph{With Search} It is possible to create a cursor at each result of a structural search. This is often part of a multi-step filtering process described in Section \ref{sec:filtering-cursors}.


\subsection{Hierarchy of Cursors}
\label{sec:hierarchy-of-cursors}
After creating multiple cursors, it is eventually necessary to switch back to a single cursor. In most text editors one of the cursors is designated as the primary cursor and there is a command to delete all other cursors. Forest instead has a command to \emph{reduce to the first\slash{}last cursor} by location in the document (remove all cursors except this one). This command interacts in a special way with Forest's novel concept of a \emph{hierarchy of cursors}.

\begin{figure}[tb]
    \includegraphics[width=\columnwidth]{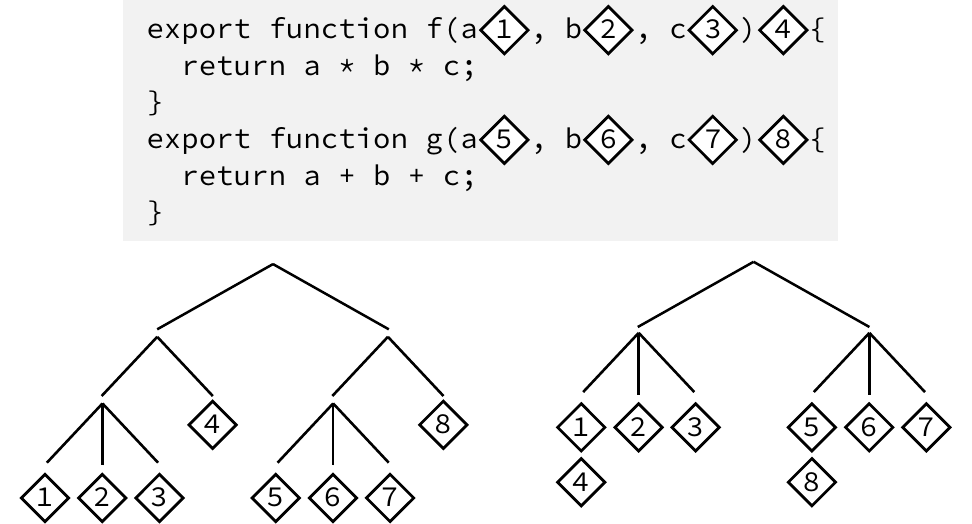}
    \captionof{figure}{Insertion locations (numbered diamonds) for type annotations shown in code (top), grouped by their location in the AST (left), and grouped by the cursor which would potentially perform the insertion in the \emph{hierarchy of cursors} (right). The hierarchy is shown assuming that locations 1 and 4 and locations 5 and 8 are edited by the same cursor.}
    \label{fig:hierarchy-example}
\end{figure}

\paragraph{Motivation} In Figure \ref{fig:hierarchy-example}, the user wants to add type annotations (e.g. \lstinline{: number}) at each location that is marked by a diamond (each function argument and the functions themselves). If they only wanted to do this for one function, they could split the cursor to have one cursor per function parameter, annotate the parameter types (locations 1 to 3), reduce to the first cursor (location 1), and finally annotate the function return type (location 4; using cursor from location 1). A natural way to perform this task for both functions is to first split the cursor to have one cursor per function, then perform the rest of the steps as for a single function. However, this does not work if \emph{reduce to first cursor} would truly leave only the first cursor, since then only the first function would get a return type annotation.

\paragraph{Principle} Forest's \emph{hierarchy of cursors} solves the issue presented above. To the best of our knowledge, this concept is novel. When a user splits a cursor into new cursors, the lineage of these cursors is tracked, effectively organizing them into a tree (right tree in Figure \ref{fig:hierarchy-example}). When the user performs an operation like \emph{reduce to first cursor}, the operation is performed \emph{per group of cursors}, where all cursors with the same parent (cursor which was split to create these cursors) are part of the same group. In Figure \ref{fig:hierarchy-example}, assuming that the cursor was split once for each function and then once for each argument (giving locations 1 to 3 and 5 to 7), \emph{reduce to first cursor} would leave the cursors on the first function argument of \emph{each} function (locations 1 and 5).

\paragraph{Multiple Levels} The hierarchy of cursors can be arbitrarily deep. The above example could be extended to split at three levels (classes, methods, and parameters). Then the first invocation of \emph{reduce to first cursor} would leave the first cursor per method and class and the second invocation would leave the first cursor per class.

\subsection{Marks}
\label{sec:marks}
While working in Forest, the user can \emph{mark} the current cursor position and later jump back to this mark. Figure \ref{fig:flatten-example} shows how marks are typically used in Forest. Forest stores marks as a selection of AST nodes and performs novel handling to ensure that marks reliably point to the same AST nodes across edits and pretty printing. Since other editors (e.g. Vim) track marks using character ranges, they lose track of marks in cases where Forest can track them reliably.


\paragraph{Motivation} Marks are useful, but not necessary with a single cursor, since it is always possible to manually move the cursor back to its old location. With multiple cursors, it is not always possible to manually move all cursors back to their old locations, since each cursor might need slightly different commands to move there, but commands are broadcast to all cursors. However, using marks, all cursors can be moved back at once, since they all require the same \emph{jump back to mark} command.

\paragraph{Persistence} Marks in Forest persist across edits and pretty-printing, always remaining on the same AST nodes. Each operation that modifies Forest's AST provides a function to adjust selections in the AST accordingly. This is applied to all marks to ensure that they point to equivalent AST nodes after each edit. This allows marks to be used extensively during editing, which is common in practice. No existing text editors or structural editors that we know of have marks that track AST nodes. 

\paragraph{Separating Cursors} Each cursor has its own set of marks. The fact that marks are part of a cursor's state is what allows cursors to have equal selections, but still be separated later.

\subsection{Filtering Cursors}
\label{sec:filtering-cursors}
\begin{figure}[tb]
    \includegraphics[width=\columnwidth]{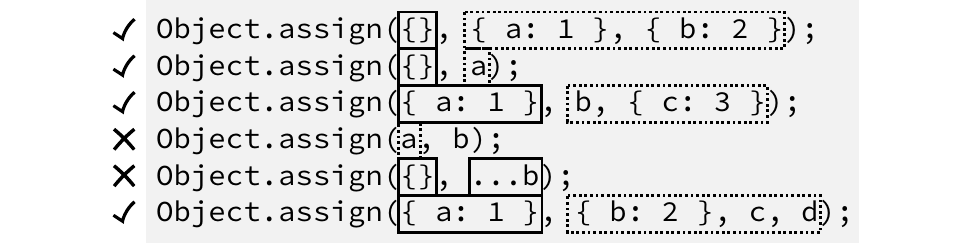}
    \captionof{figure}{Various calls to \lstinline{Object.assign} where only those with tick marks should be selected. Solid outlines are succeeding cursors and dotted outlines are failing cursors. The user creates the cursors in the left column, drops the failing ones (e.g. using drop mode), navigates to get the cursors in the right column, and keeps the failing ones (e.g. via explicit branching in strict mode). }
    \label{fig:object-assign-example}
\end{figure}

For some tasks, exactly selecting the locations to edit is complicated. In Figure \ref{fig:object-assign-example}, the user must find all calls to \lstinline{Object.assign} where the first argument is an object literal and none of the other arguments are spread elements (\lstinline{...x}). Many structural search systems cannot capture this in a single query, especially not in an easy-to-understand way.

\paragraph{Overview} In Forest, the user performs queries like this using multiple search and navigation steps in sequence. First, a structural search or direct cursor split gives all calls to \lstinline{Object.assign}. The user then navigates to the first argument (left cursors in Figure \ref{fig:object-assign-example}) and drops any cursors where this argument is not an object literal (dotted outlines). Finally, the user selects all arguments except the first and drops any calls where these arguments contain spread elements (solid outlines). The concrete mechanisms which are available to the user to perform such filtering are explained in the rest of this section.

\paragraph{Drop Mode} An implicit way to filter cursors in Forest is the \emph{multi-cursor drop mode}. After issuing a command in this mode, any \emph{failed} cursors (those which can not execute the command) are deleted. Examples of failing commands are a structural search with no results, \emph{jump to surrounding parentheses} when none exist, and \emph{reduce to first list item} when the selection is an empty list. In the \lstinline{Object.assign} example, navigating to the first argument in multi-cursor drop mode would be enough to remove any cursors where \lstinline{Object.assign} is called with no arguments, since \emph{reduce to first list item} would fail.

\paragraph{Shallow Search} Multi-cursor drop mode can also be used to handle the condition ``where the first argument is an object literal'' by using structural search. However, selecting the first argument, searching for an object literal, and dropping failed cursors does not precisely capture this condition. This method would keep cursors that contain an object literal \emph{somewhere deep inside the first argument}, for example, \lstinline$Object.assign(f({}))$. Forest has an option to use structural search to check whether the top level of the selection matches a query, without searching deeply within the selection. With this option, the check in this example is precise. If the selection matches, the command succeeds without moving the cursor. Otherwise, the command fails---which causes the cursor to be deleted in multi-cursor drop mode.

\paragraph{Limitation of Drop Mode} The main limitation of multi-cursor drop mode is that the condition which it checks cannot be inverted. For example, the condition ``none of the other arguments are spread elements'' can be expressed by \emph{keeping} all cursors where a search for spread elements \emph{fails}. This is the opposite of the behavior that multi-cursor drop mode has. Conditions like this can be expressed with Forest's \emph{explicit branching command} which is typically used in \emph{multi-cursor strict mode}.

\paragraph{Strict Mode} When a command is issued in multi-cursor strict mode, if any cursor would fail the command, then the command will not be performed at all---even for cursors where it would be possible---and each cursor will be marked succeeded or failed (visualized as solid and dotted outlines in Figure \ref{fig:object-assign-example}). Using the explicit branching command the user can now either keep the cursors that would have succeeded (solid outlines) or the ones that would have failed (dotted outlines). Keeping successful cursors is equivalent to using multi-cursor drop mode. Keeping failed cursors is only possible using this explicit branching command.

\subsection{Overlapping Cursors}
\label{sec:overlapping-cursors}
It is not immediately clear what the benefits of allowing overlapping cursors are. In this section, we will consider different kinds of overlap, show how Forest handles them, and discuss when they can be useful.

\begin{figure}[tb]
    \includegraphics[width=\columnwidth]{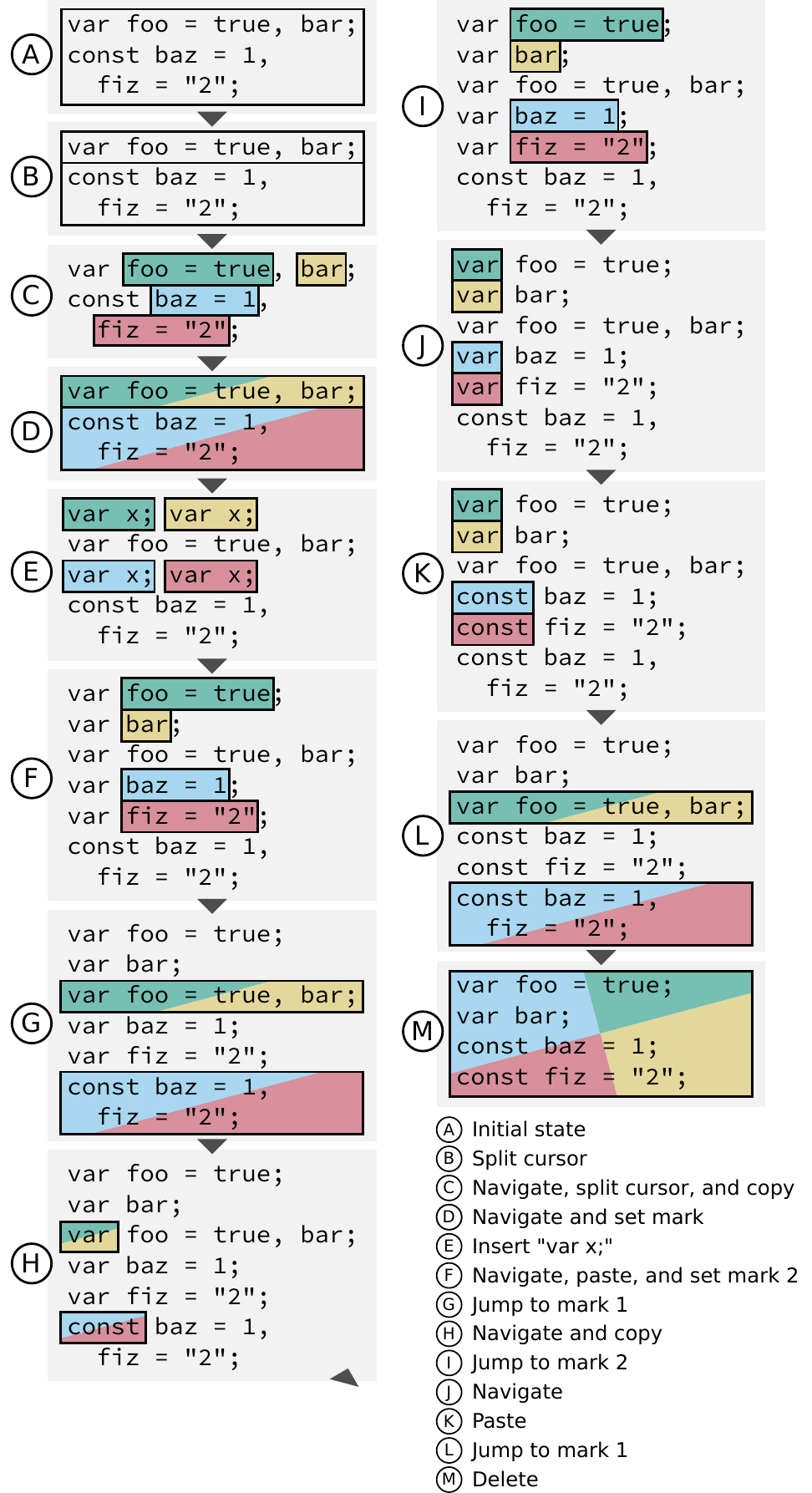}
    \captionof{figure}{An edit where variable declaration statements containing multiple declarations are flattened so that each declaration is in its own statement. This example demonstrates marks and overlapping cursors. The black outlines represent cursors. Starting from step C, each of the four cursors is marked with a different color. Rectangles with multiple colors represent overlapping cursors.}
    \label{fig:flatten-example}
\end{figure}

\paragraph{Nested Cursors} Consider the code snippet \lstinline{f(g(x))} with one cursor selecting the whole call to \lstinline{f} and another selecting the whole call to \lstinline{g}. These cursors are \emph{nested} (one strictly contains the other). This kind of overlap is discussed in Section ~\ref{sec:nested-cursors}. We call all other arrangements of cursors \emph{non-nested}.

\paragraph{Duplicate Cursors} A special case of non-nested cursors is \emph{duplicate} cursors. This is the situation where multiple cursors have the same selection range, for example, when two cursors both have the whole call to \lstinline{f} selected in \lstinline{f(g(x))}. Duplicate cursors can be separated by jumping to marks. However, duplicate cursors can also be used directly to perform insertions, in which case the typed text is inserted \emph{once for each cursor}. Duplicate cursors are ordered against each other based on the order they had before they became duplicate. This makes it predictable which instance of the newly inserted text belongs to which duplicate cursor.

\paragraph{Example with Duplicate Cursors} Both interactions with duplicate cursors (separate using marks and insert) are demonstrated in Figure \ref{fig:flatten-example}. The user creates a cursor for each variable declaration and then performs \emph{move to parent} (Step D), which results in duplicate cursors on each variable declaration list. They now perform an insertion (Step E), which creates a new statement \emph{for each duplicate cursor}, thereby also separating them. Subsequently in the example (Step H), the user has duplicate cursors on the var\slash{}const keyword, which they separate using marks.

\paragraph{Non-nested Non-duplicate Cursors} The arrangement of overlapping cursors that remains to discuss is non-nested non-duplicate cursors, as in the following example: Given the snippet \lstinline{[a, b, c]}, one cursor selects \lstinline{a, b} and one cursor selects \lstinline{b, c}. Cursors in this arrangement are generally not useful. Commands like paste (a replacement) would be ambiguous. The only non-movement command that Forest supports in this arrangement is \emph{delete}, which is done by removing a node exactly when it is contained in at least one selection.

\subsection{Nested Cursors}
\label{sec:nested-cursors}
Nested cursors are a special case of overlapping cursors. They are typically created by searching for structures that can be nested (e.g., functions or object literals). Nested cursors are uniquely challenging for multi-cursor structural editing.

\begin{figure}[tb]
    \includegraphics[width=\columnwidth]{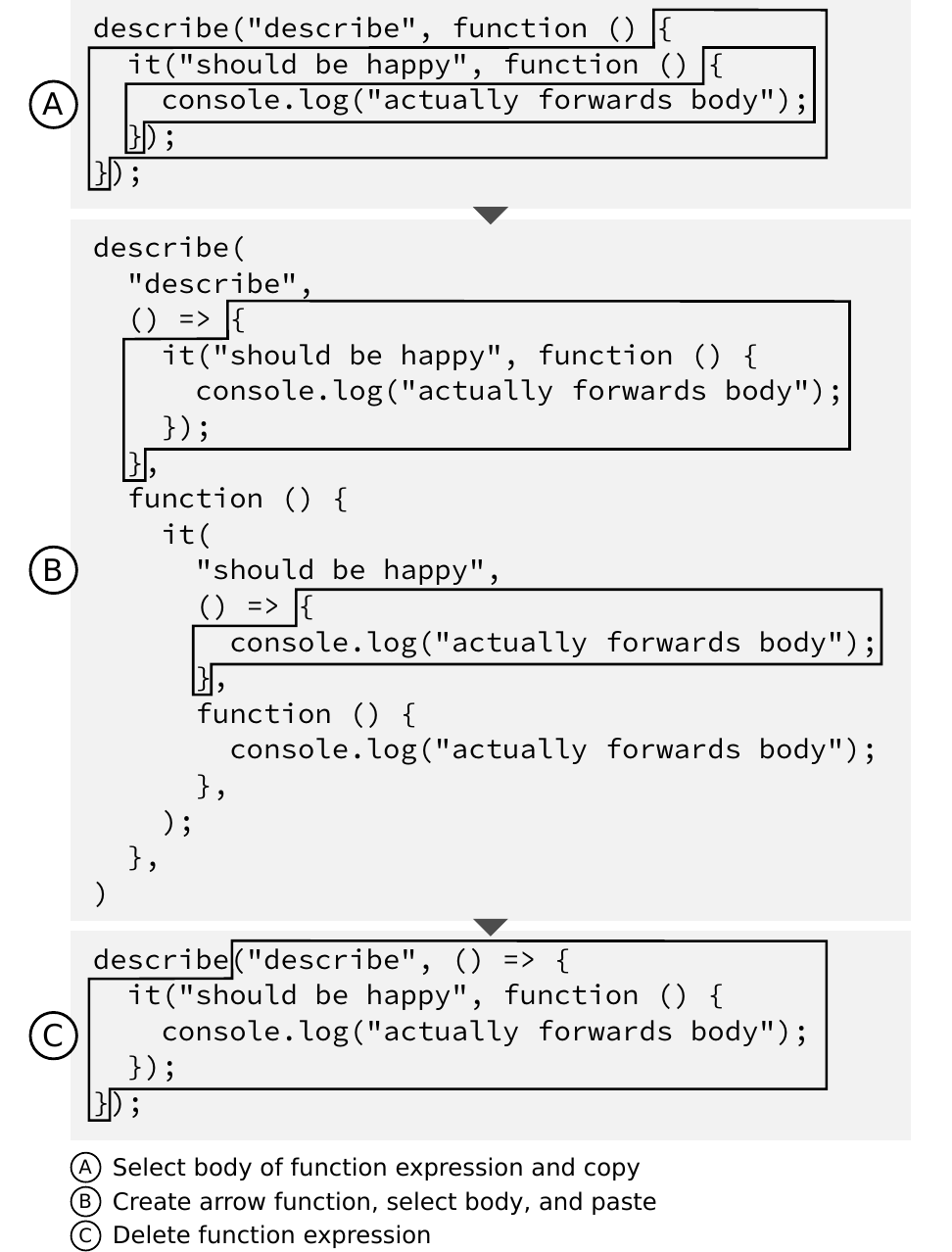}
    \captionof{figure}{An attempt at converting function expressions to arrow functions by copying and pasting the function body. The inner function is not correctly converted due to the nested copy-paste problem.}
    \label{fig:nested-copy-paste-example}
\end{figure}

\paragraph{Nested Copy-Paste Problem} Forest supports all commands (e.g., insert, paste, delete) with nested cursors. Even with this support, it is still not possible to perform certain edits with nested cursors as shown in Figure \ref{fig:nested-copy-paste-example}. To convert an individual function expression to an arrow function, the user could copy the function body, insert an arrow function, paste over its body, and delete the function expression. When editing multiple non-nested functions using multiple non-overlapping cursors, the procedure is exactly the same. However, this approach does not work with nested cursors. Since the copy command copies the function bodies before any edit operations (Step A of Figure \ref{fig:nested-copy-paste-example}), the final result is a converted version of the outer function, but all function expressions contained within are still unconverted (Step C). While the inner cursors (second cursor in Step B) do still exist and make edits, they remain in the body of the old outer function expression that eventually gets deleted. The core problem is that if a cursor copies a region containing other cursors, any future edits made by those cursors will not affect the copy.

\paragraph{Workaround} To manually work around the problem, the user could first edit with the innermost cursors, then repeat the process with the next level of surrounding cursors. This requires repeating the whole edit as many times as the deepest level of nesting (2 in our example). Forest provides a command to \emph{remove all cursors except the outermost\slash{}innermost ones}, so that at least the user does not have to manually select the cursors for each level.

%
%
%

\section{Evaluation}
\label{sec:evaluation-refactoring-scripts}
We propose multi-cursor structural editors as an interactive equivalent to refactoring scripts based on AST transformation. In order to investigate whether multi-cursor structural editors are capable of performing edits for which refactoring scripts are typically used, we collected real refactoring scripts from GitHub and attempted to perform their edits interactively in Forest.

\subsection{Method}
\paragraph{Script Framework} We considered only scripts written with the jscodeshift ~\cite{noauthor_jscodeshift_2021} framework. It is the de-facto standard for scripts that refactor JavaScript and TypeScript code. Although scripts that use the TypeScript compiler API instead of jscodeshift exist, we did not consider them, since jscodeshift is more widely used and has been available for longer. Generally, scripts that use the TypeScript compiler API are similar to those which use jscodeshift.

\paragraph{Finding Scripts} Since there is no official collection of representative jscodeshift scripts, \changed{we considered those in the most-starred list} of jscodeshift scripts on GitHub ~\cite{safronov_awesome_2021}. \changed{The original scripts are contained in our archived artifact.} There is also a longer list ~\cite{chandran_awesome_2021} which we did not use, although we expect that using it would give qualitatively similar results. \changed{We ignored 4 repositories from the list. These repositories and the reasons for ignoring them are described in appendix \ref{sec:listing-ignored-repos}. We ignored individual 12 refactoring scripts. The reasons for ignoring individual scripts (e.g. not having example code or being unrealistically simple) are given in appendix \ref{sec:listing-scripts}.}

\paragraph{Classification}
Most repositories in the list contained multiple refactoring scripts. We considered each script separately. We inspected the scripts themselves, as well as the before/after example programs that were used as test cases. Each script was classified based on whether we could reproduce its edits in Forest using a best-effort approach:

\begin{itemize}
    \item \emph{No}: From reading the code it is clear that this or a similar refactor would not be possible in Forest for a major reason.
    \item \emph{Maybe}: This refactor or something similar would likely be possible, but would require a new feature, extra manual work, or a trade-off in the result.
    \item \emph{Yes}: This refactor is clearly possible. The same kinds of limitations as for \emph{Maybe} are acceptable, but they must be quite minor.
\end{itemize}

\paragraph{Editing Attempts} For scripts that were classified as \emph{Maybe} or \emph{Yes}, we tried performing their edits in Forest. \changed{Some of these edits can be viewed step-by-step in Forest (Footnote \ref{footnote:forest-demo-site}).} We wrote down any unforeseen issues and limitations of our solution. Since Forest is a prototype, it lacks support for certain language constructs. We approximated unsupported constructs using existing supported syntax. For example, \linebreak\lstinline$import { render } from "react-dom"$ was written as \linebreak\lstinline$fakeImport([render], "react-dom")$. These cases still count towards the ``unsupported syntax'' issue in our results.

\paragraph{Previously Used Examples} 
Prior to conducting this evaluation, we had already tried to reproduce the edits of some refactoring scripts in Forest and added new features accordingly. Three of the scripts \cite{nakazawa_cpojerjs-codemod_2022} included in our evaluation were previously used. Each of these scripts has a corresponding example (Figures \ref{fig:flatten-example}, \ref{fig:object-assign-example}, and \ref{fig:nested-copy-paste-example}).

\subsection{Results}
\label{sec:results-refactoring-scripts}
We used a total of 48 refactoring scripts in our evaluation \changed{(not counting ignored scripts)}. We classified them as follows: 20 \emph{No}, 17 \emph{Maybe}, and 11 \emph{Yes}. Table \ref{table:refactoring-script-issues} lists the issues that we encountered during our editing attempts. The rest of this section describes those issues. We focus on commonly occurring issues that are not clear from the name alone. Additionally, we describe some less common issues that we consider especially important. \changed{Appendix \ref{sec:listing-scripts} contains a detailed listing which describes the functionality of each script and gives our evaluation result, including a list of encountered issues.}

\begin{table*}[ht]
\caption{Issues encountered while performing the edits of real-world refactoring scripts (Section \ref{sec:evaluation-refactoring-scripts}). The numbers in each row indicate the number of scripts where we encountered the given issue. The \emph{Total} column counts scripts regardless of how we classified them. The other columns only count scripts that had the corresponding classification. For example, we encountered the issue ``Have to recreate cursors multiple times'' with 8 scripts, 2 of which were classified \emph{No}. Note that each script may have multiple issues. Scripts classified \emph{Yes} often had no issues. The issues (rows in \emph{italics}) are grouped into categories (rows in \textbf{bold}). The ellipsis ($\cdots$) indicates that some issues were omitted for brevity. See Table \ref{table:refactoring-script-issues-full} for a full table.}
\label{table:refactoring-script-issues}
\begin{tabular}{@{}ccccl@{}} \toprule
Total & No & Maybe & Yes & \textbf{Category} or \emph{Specific Issue} \\\midrule
\textbf{27} & \textbf{7} & \textbf{17} & \textbf{3} & \textbf{Missing features that require conceptual changes} \\
\emph{8} & \emph{2} & \emph{5} & \emph{1} & \emph{Have to recreate cursors multiple times} \\
\emph{6} & \emph{2} & \emph{4} & \emph{0} & \emph{Cannot handle separately found locations together} \\
\emph{2} & \emph{1} & \emph{1} & \emph{0} & \emph{None-one-many issue} \\
\multicolumn{5}{c}{$\cdots$} \\
\hline
\textbf{21} & \textbf{12} & \textbf{9} & \textbf{0} & \textbf{Missing features that do not require conceptual changes} \\
\emph{5} & \emph{2} & \emph{3} & \emph{0} & \emph{No strict ``find usages of variable''} \\
\emph{2} & \emph{2} & \emph{0} & \emph{0} & \emph{Cannot remove duplicate items} \\
\multicolumn{5}{c}{$\cdots$} \\
\hline
\textbf{11} & \textbf{11} & \textbf{0} & \textbf{0} & \textbf{Edit is too complicated for multi-cursor structural editing} \\
\emph{3} & \emph{3} & \emph{0} & \emph{0} & \emph{Lookup tables are possible but impractical} \\
\multicolumn{5}{c}{$\cdots$} \\
\hline
\textbf{11} & \textbf{6} & \textbf{4} & \textbf{1} & \textbf{Unsupported syntax} \\
\emph{11} & \emph{6} & \emph{4} & \emph{1} & \emph{Unsupported syntax} \\
\bottomrule
\vspace{0.1cm}
\end{tabular}
\end{table*}

\paragraph{``Unsupported syntax''} The most common issue in our evaluation was ``unsupported syntax''. However, this was almost never the issue that caused a script to be classified \emph{No} --- it just happened to be a common issue overall. The only exception was a script that required template literals, which are not similar to any supported syntax in Forest and would require extensive work to accommodate.

\paragraph{``Nested copy-paste would be an issue''} This issue is discussed extensively in Section \ref{sec:nested-cursors}, since it was known to us before this evaluation.

\paragraph{``Have to recreate cursor multiple times''} Consider the cursors on the left side of Figure \ref{fig:object-assign-example}. It is possible to keep only the cursors containing object literals and perform an edit with them. It is also possible to keep only the cursors containing identifiers and perform a different edit with them. However, it is not possible to perform these two edits in sequence, without manually recreating some cursors. Once all cursors except those containing object literals have been deleted, there is no command to restore them (``undo selection change'' does not work across edits; marks are saved per cursor, so they cannot recreate deleted cursors). By deleting cursors before an edit, Forest can represent edits with pseudocode of the form:\\
\begin{minipage}{\columnwidth}
\begin{lstlisting}
if (cursor.matchesCondition(conditionA)) {
    cursor.performEdit(editA)
}
// no further edits
\end{lstlisting}
\end{minipage}
However, since checking the condition is done by deleting cursors, edits of the following form can not be represented: 
\begin{minipage}{\columnwidth}
\begin{lstlisting}
if (cursor.matchesCondition(conditionA)) {
  cursor.performEdit(editA)  
}
cursor.performEdit(editB)
\end{lstlisting}
\end{minipage}

\pagebreak
\paragraph{``Cannot handle separately found locations together''} Consider a program containing multiple object literals, each with a similar set of fields. A user would like to edit some of these object literals simultaneously using multiple cursors. Specifically, they would like to edit object literals which are arguments in a call to the function \lstinline{f} (\lstinline$f({ ... })$), as well as object literals used in return statements (\lstinline$return { ... }$). In Forest, the user can find object literals used in a call to \lstinline{f}, edit those, then delete all cursors except one, find all object literals used in return statements, and edit those. However, there is no way for the user to find object literals used \emph{either} in a call to \lstinline{f} \emph{or} in a return statement, then edit all of them simultaneously (regardless of which of the two conditions they matched). The following pseudocode describes such an edit (which is not possible in Forest):
\begin{minipage}{\columnwidth}
\begin{lstlisting}
if (
  cursor.matchesCondition(conditionA) ||
  cursor.matchesCondition(conditionB)
) {
  cursor.performEdit(editA)
}
\end{lstlisting}
\end{minipage}


\paragraph{``None-one-many issue''} Some commands in Forest work differently depending on whether a list contains no items, one item, or more than one item. Consider three cursors focused on the argument lists (excluding parentheses) of \lstinline{a()}, \lstinline{b(x.y)} and \lstinline{c(x.y, z)}. If the user executes the ``split cursor'' command (in relaxed multi cursor mode), they would get one cursor in the argument list of \lstinline{a} (unchanged, since the command failed), a cursor on \lstinline{x} in \lstinline{b}, a cursor on \lstinline{y} in \lstinline{b}, a cursor on \lstinline{x.y} in \lstinline{c}, and a cursor on \lstinline{z} in \lstinline{c}. The differences between one item and more than one item are typically caused by Forest's handling of equivalent cursors. The differences between no items and some items are typically caused by commands requiring at least one item to function.


\section{Related Work}
\label{sec:related-work}
This section gives an overview of existing structural code editors, as well as existing approaches for performing repetitive code edits which were shown in Figure \ref{fig:interactive-textual-structural-space}. 

\paragraph{Early Structural Editors} \emph{Structural code editors} (also called \emph{structure editors} or \emph{projectional editors}) are editors in which the navigation and editing commands operate on a tree representation of a program, rather than allowing the user to arbitrarily modify the text of a source file. Structural editors have a long history, with some created as early as the 1970s (Emily ~\cite{hansen_user_1971}), and many more created around the 1980s (Mentor ~\cite{donzeau-gouge_programming_1980}, Cornell program synthesizer ~\cite{teitelbaum_cornell_1981}, GANDALF ~\cite{ellison_evolution_1985}, Syned ~\cite{gansner_syned_1983}, Lispedit ~\cite{mikelsons_interactive_1983}, Poe ~\cite{fischer_poe_1984}). These projects made some assumptions that no longer hold in today's environment. Many older structural editors were designed for Pascal, which has a simpler syntax compared to TypeScript and other modern languages. We believe that complex syntax makes some designs impractical, such as dropdown menus to select the type of AST node to insert, or hints showing possible locations for optional children, because the user may be overwhelmed by the number of choices and hints. Designing for acceptable performance was a major constraint for old editors, but today's more powerful computers require much less focus on performance and make new designs possible. Additionally, many features that motivated old structural editors are now commonplace in our structure-aware text editors. Some examples are scope-aware auto-completion, jump-to-definition, refactors for extracting and moving code, and continuous type checking.

\paragraph{Newer Structural Editors} Since the 1980s there has been less active development of structural editors. Some notable newer projects are MPS ~\cite{noauthor_mps_nodate} (usually Java as base language; language workbench), Envision ~\cite{asenov_envision_2014} (a subset of Java; focus on visualization), Hazel ~\cite{omar_live_2019} (custom functional language), Lamdu ~\cite{chuchem_lamdu_nodate} (custom functional language), and GopCaml ~\cite{gopinathan_gopcaml_2021} (OCaml; plugin for Emacs). MPS is by far the most widely used structural editor, yet it is almost completely unknown compared to text editors. The fact that MPS is a language workbench can add complexity for users, for example when working with variable references ~\cite{berger_efficiency_2016}. GopCaml is the closest to our work in that it fully supports OCaml, which is a widely known language with complex syntax.

\paragraph{Multi-cursor Editing}
Multi-cursor editing (also called multiple selection or simultaneous editing) is a feature in text editors which lets the user create more than one cursor/selection in a document. In response to a user's command (e.g., pressing backspace) every cursor moves and edits simultaneously. It was first described by Miller ~\cite{miller_interactive_2001}. Sublime Text (2008) seems to be the first widely used editor with a multi-cursor feature. Since its release, this has become a standard feature in most common code editors (e.g., Ace, VScode, emacs with a plugin, IntelliJ, and Notepad++). However, no designs for multi-cursor structural code editors have been proposed. We believe that multi-cursor editing and structural editing have not been investigated in combination, because multi-cursor editing became widespread much later than the peak of interest in structural editors. Note that although GopCaml ~\cite{gopinathan_gopcaml_2021} can be combined with multiple-cursors for Emacs ~\cite{sveen_multiple-cursors_2021}, there is almost no special handling or discussion for this combination.


\paragraph{Kakoune}
Kakoune~\cite{coste_kakoune_nodate} is a unique editor which combines multi-cursor as a central primitive with a large set of text manipulation commands. It has some similarities to structural editors while still being a text editor. In Kakoune, the user always has at least one selection. Selections are always a text range, not a text location. The user can widen, narrow or split the selection, often using regular expressions. This can be used to navigate in a similar way to moving up and down a tree in a structural editor. Although, since Kakoune itself has no understanding of programming language syntax, it is not a structural editor. The rich set of commands in Kakoune allows much more complex edits than in other multi-cursor text editors. By issuing multiple commands that include regular expressions, a Kakoune user can achieve similar edits to structural regular expressions \cite{pike_structural_1987}. 

\paragraph{Macros}
Macros are either recordings of keypresses that trigger editor commands or programs in a scripting language that can control the editor. They can be used to achieve similar results to multi-cursor editing. Although multi-cursor editing is relatively new, macros are a feature even in older text editors like Emacs and Vi. Some structural editors have macro support (e.g., Mentor procedures ~\cite{melese_mentor-v5_1985}). The main advantage of multi-cursor editing compared to recorded macros is that multi-cursor editing shows the effect of every edit in every location simultaneously, but a recorded macro only shows edits in one location while recording.

\paragraph{AST Transformation Scripts}
To perform precise large-scale code changes it is possible to write scripts that operate on abstract syntax trees. This can either be done using specialized transformation languages such as TXL ~\cite{cordy_txl_2006}, or using general-purpose languages with compiler libraries for parsing and printing. JavaScript developers generally use jscodeshift ~\cite{noauthor_jscodeshift_2021} or the TypeScript compiler API for creating such scripts. An example of such transformation scripts are the React codemod scripts ~\cite{noauthor_reactjsreact-codemod_2021}. Such AST transformation scripts are a heavyweight approach --- they require specialized knowledge and are not created in an interactive editor. We believe that writing such scripts is tedious, and may take longer than performing the edits manually if the number of edited locations is small.

\paragraph{Programming by Demonstration}
Programming by demonstration tools are another option for performing repetitive code changes. The user performs the desired change in a few locations and the system finds more similar changes and recommends equivalent edits. Examples of such systems are Sydit ~\cite{meng_systematic_2011} and reCode ~\cite{ni_recode_2021}. Programming by demonstration systems abstract edits from concrete examples, while in multi-cursor systems the user directly performs edits with sufficiently abstract commands.

\paragraph{Scripts vs Editors}
To make large scale repetitive changes to text files, a programmer may choose to write a small program to perform the changes. Such text editing scripts are often written in scripting languages (e.g., Bash or Python) and contain operations like regular expression searches, string splitting, or loops. However, for most editing tasks, programmers will perform these edits in a normal text editor instead of writing scripts. Some text editors (e.g., vim) can perform many of the operations used in text editing scripts, especially when combined with macros. Such text editors could be seen as the interactive equivalent of text editing scripts. Analogously, we believe that multi-cursor structural editors are the interactive equivalent of AST transformation scripts. This idea is investigated in Section \ref{sec:evaluation-refactoring-scripts}.

\section{Discussion}
\label{sec:discussion}
In Section \ref{sec:evaluation-refactoring-scripts}, we successfully used Forest to perform the edits from 11 of the 48 refactoring scripts which we tested. However, during this process, we discovered various issues that limited the possible edits of 17 further scripts. Most of these issues are related to multi-cursor editing.

\changed{\paragraph{Current Scope of Forest} In our evaluation, we found that Forest can be used to perform a variety of common edits. Examples of such edits are \emph{``swap two arguments in all calls to a function with a specific name''} or \emph{``replace selected function calls by their first argument''}. The various features of Forest can often be composed to achieve complex edits. Consider the edit \emph{``move the selected property to the start of the containing object literal''}, which is possible in Forest. By performing that edit in conjunction with Forest's multi-cursor and structural search features, the edit \emph{``in every object literal containing a specific property, move that property to the start of the object literal''} also becomes possible. Further examples of complex edits achieved by combining Forest's features are: \emph{``add an extra argument to every function call whose name matches a regular expression''}, \emph{``add a type annotation to every variable declaration which does not have a type annotation, is contained in a specific function, and whose name matches a regular expression''}, \emph{``find all array literals which are assigned as values in \texttt{const} variable declarations''}, and \emph{``add a print at the start of every function that contains \texttt{throw} statements''}.}

\changed{\paragraph{Future Scope of Forest} Our evaluation highlighted multiple limitations in the current design of Forest. In order to expand the set refactorings that are possible in Forest, we plan to address the issues listed under ``Missing features that do not require conceptual changes'' in Table \ref{table:refactoring-script-issues-full}. The refactorings that this makes possible are generally stricter or larger-scale versions of the currently supported ones. Examples are \emph{``wrap every usage of a specific variable in a function call''} and \emph{``add a type annotation to every variable declaration whose variable is used as the first argument in calls to a specific function''} (currently not possible since variables can be found by name, but not by strict reference), as well as \emph{``add an extra argument to every function call where the called function is declared in a specific file''} (requires both following references and working with multiple files). We believe that many of the issues listed under \emph{``Missing features that require conceptual changes''} could also be addressed, although their design will require careful consideration.}

\changed{\paragraph{Out-of-Scope Edits} We believe that certain edits will never be performed in a multi-cursor structural editor such as Forest. For example, refactorings where many special cases must be handled are not practical to describe by interactively issuing commands. Any edits with specialized logic, such as dead code detection or parsing string literals, are not possible in Forest, since they would require a general purpose programming interface. However, such tasks are handled well by refactoring scripts. Additionally, since Forest works with an abstract syntax tree representation of code, it can not be used to adjust how code is formatted as text or to work with files containing syntax errors.}

\changed{\paragraph{Adoption} We envision Forest being adopted by users of existing specialized developer tools, such as multi-cursor editing, command based editors such as Vim, advanced refactorings in IDEs, or refactoring scripts. However, we believe that Forest will only be adopted by a narrow expert audience, since many developers do not use any of the previously listed tools, which have comparable complexity to Forest. Understanding how difficult it is for developers to learn to use a multi-cursor structural editor like Forest productively is an interesting direction for future study. Other than the steep learning curve, the are two major limitations that we expect in a production version of Forest. First, non-standard syntax is not allowed in source files. Second, all edited files must be automatically pretty-printed, since Forest provides no control over code formatting. Many projects already enforce automatic pretty printing in their codebase.}

\paragraph{\changed{Combining with Refactoring Scripts}} Based on our evaluation in Section \ref{sec:evaluation-refactoring-scripts}, we believe that multi-cursor structural editing could become a viable alternative to writing refactoring scripts \emph{in simple cases}, but that such editors cannot replace refactoring scripts in general. Based on our usage of Forest, we think that interactivity is a major strength of multi-cursor editing. It may be possible to combine the expressiveness of refactoring scripts with the interactivity of multi-cursor editing. For example, consider a hypothetical IDE where the user writes refactoring scripts, but can interactively inspect intermediate search results and transformations. By treating the intermediate search results as cursors in a multi-cursor editor, the user would be able to define transformations by performing them interactively. We think that this is an interesting direction for future research.

\paragraph{Relationship to Textual Multi-cursor}
Multi-cursor editing in a structural editor is not fundamentally different than multi-cursor editing in a text editor, but it is arguably more powerful, because the power of multi-cursor editing \changed{depends on the available single cursor editing commands.}

Consider a minimal text editor where the only operations are \emph{move cursor by one character}, \emph{delete character}, and \emph{insert character}. It would generally only be possible to edit two code snippets using multi-cursor in such an editor if the code had the same structure, the code was formatted identically, and matching literals had the same length. Since most text editors extend this minimal model with commands like \emph{move cursor by one word} and \emph{move cursor to end of line}, matching literals generally do not need to have the same length and minor code formatting differences can be tolerated. Since structural editors add structural movement and editing commands, they can tolerate most formatting differences.

\addtolength{\textheight}{-3.02cm}

\section{Conclusion}
\label{sec:conclusion}
We have presented Forest, an editor with unique integration between structural editing and multi-cursor editing. We have introduced the novel concept of a hierarchy of cursors, which allows multi-cursor edits to be reused as part of other multi-cursor edits. Additionally, we have designed an interactive filtering system by introducing alternative multi-cursor modes.
We attempted to perform edits from real-world refactoring scripts, and showed that Forest could perform the edits comparably well to the scripts in 11 of the 48 cases. With improvements to our prototype, we expect that a further 17 refactoring scripts from our evaluation to be replaced. We believe that multi-cursor structural editors could be useful for performing a wide range of specialized ad-hoc refactorings.

\begin{acks}
This research was supported by a Ministry of Education (MOE) Academic Research Fund (AcRF) Tier 1 grant.
\end{acks}

\bibliographystyle{ACM-Reference-Format}
\bibliography{main}


\begin{thebibliography}{28}


\ifx \showCODEN    \undefined \def \showCODEN     #1{\unskip}     \fi
\ifx \showDOI      \undefined \def \showDOI       #1{#1}\fi
\ifx \showISBNx    \undefined \def \showISBNx     #1{\unskip}     \fi
\ifx \showISBNxiii \undefined \def \showISBNxiii  #1{\unskip}     \fi
\ifx \showISSN     \undefined \def \showISSN      #1{\unskip}     \fi
\ifx \showLCCN     \undefined \def \showLCCN      #1{\unskip}     \fi
\ifx \shownote     \undefined \def \shownote      #1{#1}          \fi
\ifx \showarticletitle \undefined \def \showarticletitle #1{#1}   \fi
\ifx \showURL      \undefined \def \showURL       {\relax}        \fi
\providecommand\bibfield[2]{#2}
\providecommand\bibinfo[2]{#2}
\providecommand\natexlab[1]{#1}
\providecommand\showeprint[2][]{arXiv:#2}

\bibitem[\protect\citeauthoryear{??}{ase}{[n.d.]}]%
        {asenov_envision_2014}
 \bibinfo{year}{[n.d.]}\natexlab{}.
\newblock \showarticletitle{Envision: {A} fast and flexible visual code editor
  with fluid interactions ({Overview})}. \bibinfo{address}{Melbourne,
  Australia}.
\newblock


\bibitem[\protect\citeauthoryear{??}{noa}{[n.d.]}]%
        {noauthor_mps_nodate}
 \bibinfo{year}{[n.d.]}\natexlab{}.
\newblock \bibinfo{title}{{MPS}: {The} {Domain}-{Specific} {Language} {Creator}
  by {JetBrains}}.
\newblock
\newblock
\urldef\tempurl%
\url{https://www.jetbrains.com/mps/}
\showURL{%
\tempurl}


\bibitem[\protect\citeauthoryear{??}{noa}{2021a}]%
        {noauthor_jscodeshift_2021}
 \bibinfo{year}{2021}\natexlab{a}.
\newblock \bibinfo{title}{jscodeshift}.
\newblock
\newblock
\urldef\tempurl%
\url{https://github.com/facebook/jscodeshift}
\showURL{%
\tempurl}
\newblock
\shownote{original-date: 2015-03-07T00:32:16Z.}


\bibitem[\protect\citeauthoryear{??}{noa}{2021b}]%
        {noauthor_reactjsreact-codemod_2021}
 \bibinfo{year}{2021}\natexlab{b}.
\newblock \bibinfo{title}{reactjs/react-codemod}.
\newblock
\newblock
\urldef\tempurl%
\url{https://github.com/reactjs/react-codemod}
\showURL{%
\tempurl}
\newblock
\shownote{original-date: 2015-10-19T20:47:22Z.}


\bibitem[\protect\citeauthoryear{Berger, Völter, Jensen, Dangprasert, and
  Siegmund}{Berger et~al\mbox{.}}{2016}]%
        {berger_efficiency_2016}
\bibfield{author}{\bibinfo{person}{Thorsten Berger}, \bibinfo{person}{Markus
  Völter}, \bibinfo{person}{Hans~Peter Jensen}, \bibinfo{person}{Taweesap
  Dangprasert}, {and} \bibinfo{person}{Janet Siegmund}.}
  \bibinfo{year}{2016}\natexlab{}.
\newblock \showarticletitle{Efficiency of projectional editing: a controlled
  experiment}. In \bibinfo{booktitle}{\emph{Proceedings of the 2016 24th {ACM}
  {SIGSOFT} {International} {Symposium} on {Foundations} of {Software}
  {Engineering}}} \emph{(\bibinfo{series}{{FSE} 2016})}.
  \bibinfo{publisher}{Association for Computing Machinery},
  \bibinfo{address}{New York, NY, USA}, \bibinfo{pages}{763--774}.
\newblock
\showISBNx{978-1-4503-4218-6}
\urldef\tempurl%
\url{https://doi.org/10.1145/2950290.2950315}
\showDOI{\tempurl}


\bibitem[\protect\citeauthoryear{Chandran}{Chandran}{2021}]%
        {chandran_awesome_2021}
\bibfield{author}{\bibinfo{person}{Rajasegar Chandran}.}
  \bibinfo{year}{2021}\natexlab{}.
\newblock \bibinfo{title}{Awesome {Codemods}}.
\newblock
\newblock
\urldef\tempurl%
\url{https://github.com/rajasegar/awesome-codemods}
\showURL{%
\tempurl}
\newblock
\shownote{original-date: 2019-12-11T00:38:56Z.}


\bibitem[\protect\citeauthoryear{Chuchem and Lotem}{Chuchem and
  Lotem}{[n.d.]}]%
        {chuchem_lamdu_nodate}
\bibfield{author}{\bibinfo{person}{Yair Chuchem} {and} \bibinfo{person}{Eyal
  Lotem}.} \bibinfo{year}{[n.d.]}\natexlab{}.
\newblock \bibinfo{title}{Lamdu}.
\newblock
\newblock
\urldef\tempurl%
\url{https://www.lamdu.org/}
\showURL{%
\tempurl}


\bibitem[\protect\citeauthoryear{Cordy}{Cordy}{2006}]%
        {cordy_txl_2006}
\bibfield{author}{\bibinfo{person}{James~R. Cordy}.}
  \bibinfo{year}{2006}\natexlab{}.
\newblock \showarticletitle{The {TXL} source transformation language}.
\newblock \bibinfo{journal}{\emph{Science of Computer Programming}}
  \bibinfo{volume}{61}, \bibinfo{number}{3} (\bibinfo{date}{Aug.}
  \bibinfo{year}{2006}), \bibinfo{pages}{190--210}.
\newblock
\showISSN{01676423}
\urldef\tempurl%
\url{https://doi.org/10.1016/j.scico.2006.04.002}
\showDOI{\tempurl}


\bibitem[\protect\citeauthoryear{Coste}{Coste}{[n.d.]}]%
        {coste_kakoune_nodate}
\bibfield{author}{\bibinfo{person}{Maxime Coste}.}
  \bibinfo{year}{[n.d.]}\natexlab{}.
\newblock \bibinfo{title}{Kakoune - {Official} site}.
\newblock
\newblock
\urldef\tempurl%
\url{https://kakoune.org/}
\showURL{%
\tempurl}


\bibitem[\protect\citeauthoryear{Donzeau-Gouge, Huet, Lang, and
  Kahn}{Donzeau-Gouge et~al\mbox{.}}{1980}]%
        {donzeau-gouge_programming_1980}
\bibfield{author}{\bibinfo{person}{Véronique Donzeau-Gouge},
  \bibinfo{person}{Gérard Huet}, \bibinfo{person}{Bernard Lang}, {and}
  \bibinfo{person}{Gilles Kahn}.} \bibinfo{year}{1980}\natexlab{}.
\newblock \bibinfo{booktitle}{\emph{Programming environments based on
  structured editors: the {Mentor} experience}}.
\newblock \bibinfo{type}{{T}echnical {R}eport}. \bibinfo{institution}{INRIA}.
\newblock


\bibitem[\protect\citeauthoryear{Ellison and Staudt}{Ellison and
  Staudt}{1985}]%
        {ellison_evolution_1985}
\bibfield{author}{\bibinfo{person}{Robert~J. Ellison} {and}
  \bibinfo{person}{Barbara~J. Staudt}.} \bibinfo{year}{1985}\natexlab{}.
\newblock \showarticletitle{The evolution of the {GANDALF} system}.
\newblock \bibinfo{journal}{\emph{Journal of Systems and Software}}
  \bibinfo{volume}{5}, \bibinfo{number}{2} (\bibinfo{date}{May}
  \bibinfo{year}{1985}), \bibinfo{pages}{107--119}.
\newblock
\showISSN{0164-1212}
\urldef\tempurl%
\url{https://doi.org/10.1016/0164-1212(85)90012-3}
\showDOI{\tempurl}


\bibitem[\protect\citeauthoryear{Fischer, Johnson, Mauney, Pal, and
  Stock}{Fischer et~al\mbox{.}}{1984}]%
        {fischer_poe_1984}
\bibfield{author}{\bibinfo{person}{C.~N. Fischer}, \bibinfo{person}{Gregory~F.
  Johnson}, \bibinfo{person}{Jon Mauney}, \bibinfo{person}{Anil Pal}, {and}
  \bibinfo{person}{Daniel~L. Stock}.} \bibinfo{year}{1984}\natexlab{}.
\newblock \showarticletitle{The {Poe} language-based editor project}. In
  \bibinfo{booktitle}{\emph{Proceedings of the first {ACM} {SIGSOFT}/{SIGPLAN}
  software engineering symposium on {Practical} software development
  environments}} \emph{(\bibinfo{series}{{SDE} 1})}.
  \bibinfo{publisher}{Association for Computing Machinery},
  \bibinfo{address}{New York, NY, USA}, \bibinfo{pages}{21--29}.
\newblock
\showISBNx{978-0-89791-131-3}
\urldef\tempurl%
\url{https://doi.org/10.1145/800020.808245}
\showDOI{\tempurl}


\bibitem[\protect\citeauthoryear{Gansner, Horgan, Moore, Surko, Swartwout, and
  Reppy}{Gansner et~al\mbox{.}}{1983}]%
        {gansner_syned_1983}
\bibfield{author}{\bibinfo{person}{E. Gansner}, \bibinfo{person}{J.~R. Horgan},
  \bibinfo{person}{D.~J. Moore}, \bibinfo{person}{P. Surko},
  \bibinfo{person}{D. Swartwout}, {and} \bibinfo{person}{J. Reppy}.}
  \bibinfo{year}{1983}\natexlab{}.
\newblock \bibinfo{booktitle}{\emph{Syned -- {A} {Language}-{Based} {Editor}
  for an {Interactive} {Programming} {Environment}}}.
\newblock \bibinfo{type}{{T}echnical {R}eport}.
\newblock


\bibitem[\protect\citeauthoryear{Gopinathan}{Gopinathan}{2021}]%
        {gopinathan_gopcaml_2021}
\bibfield{author}{\bibinfo{person}{Kiran Gopinathan}.}
  \bibinfo{year}{2021}\natexlab{}.
\newblock \bibinfo{title}{{GopCaml}: {A} {Structural} {Editor} for {OCaml}}.
\newblock
\newblock
\urldef\tempurl%
\url{https://icfp21.sigplan.org/details/ocaml-2021-papers/11/GopCaml-A-Structural-Editor-for-OCaml}
\showURL{%
\tempurl}


\bibitem[\protect\citeauthoryear{Hansen}{Hansen}{1971}]%
        {hansen_user_1971}
\bibfield{author}{\bibinfo{person}{Wilfred~J. Hansen}.}
  \bibinfo{year}{1971}\natexlab{}.
\newblock \showarticletitle{User engineering principles for interactive
  systems}. In \bibinfo{booktitle}{\emph{Proceedings of the {May} 16-18, 1972,
  spring joint computer conference on - {AFIPS} '72 ({Spring})}}.
  \bibinfo{publisher}{ACM Press}, \bibinfo{address}{Atlantic City, New Jersey},
  \bibinfo{pages}{523}.
\newblock
\urldef\tempurl%
\url{https://doi.org/10.1145/1479064.1479159}
\showDOI{\tempurl}


\bibitem[\protect\citeauthoryear{Kim and Meng}{Kim and Meng}{2014}]%
        {kim_recommending_2014}
\bibfield{author}{\bibinfo{person}{Miryung Kim} {and} \bibinfo{person}{Na
  Meng}.} \bibinfo{year}{2014}\natexlab{}.
\newblock \showarticletitle{Recommending {Program} {Transformations}}.
\newblock In \bibinfo{booktitle}{\emph{Recommendation {Systems} in {Software}
  {Engineering}}}, \bibfield{editor}{\bibinfo{person}{Martin~P. Robillard},
  \bibinfo{person}{Walid Maalej}, \bibinfo{person}{Robert~J. Walker}, {and}
  \bibinfo{person}{Thomas Zimmermann}} (Eds.). \bibinfo{publisher}{Springer},
  \bibinfo{address}{Berlin, Heidelberg}, \bibinfo{pages}{421--453}.
\newblock
\showISBNx{978-3-642-45135-5}
\urldef\tempurl%
\url{https://doi.org/10.1007/978-3-642-45135-5_16}
\showDOI{\tempurl}


\bibitem[\protect\citeauthoryear{Meertens, Pemberton, and Rossum}{Meertens
  et~al\mbox{.}}{1992}]%
        {meertens_abc_1992}
\bibfield{author}{\bibinfo{person}{L. Meertens}, \bibinfo{person}{S.
  Pemberton}, {and} \bibinfo{person}{G. Rossum}.}
  \bibinfo{year}{1992}\natexlab{}.
\newblock \bibinfo{booktitle}{\emph{The {ABC} structure editor --
  {Structure}-based editing for the {ABC} programming environment.}}
\newblock \bibinfo{type}{{T}echnical {R}eport}.
\newblock


\bibitem[\protect\citeauthoryear{Melese, Migot, and Verove}{Melese
  et~al\mbox{.}}{1985}]%
        {melese_mentor-v5_1985}
\bibfield{author}{\bibinfo{person}{B Melese}, \bibinfo{person}{V Migot}, {and}
  \bibinfo{person}{D Verove}.} \bibinfo{year}{1985}\natexlab{}.
\newblock \bibinfo{booktitle}{\emph{The {Mentor}-{V5} documentation}}.
\newblock \bibinfo{type}{{T}echnical {R}eport}. \bibinfo{institution}{INRIA}.
\newblock


\bibitem[\protect\citeauthoryear{Meng, Kim, and McKinley}{Meng
  et~al\mbox{.}}{2011}]%
        {meng_systematic_2011}
\bibfield{author}{\bibinfo{person}{Na Meng}, \bibinfo{person}{Miryung Kim},
  {and} \bibinfo{person}{Kathryn~S. McKinley}.}
  \bibinfo{year}{2011}\natexlab{}.
\newblock \showarticletitle{Systematic editing: generating program
  transformations from an example}. In \bibinfo{booktitle}{\emph{Proceedings of
  the 32nd {ACM} {SIGPLAN} {Conference} on {Programming} {Language} {Design}
  and {Implementation}}} \emph{(\bibinfo{series}{{PLDI} '11})}.
  \bibinfo{publisher}{Association for Computing Machinery},
  \bibinfo{address}{New York, NY, USA}, \bibinfo{pages}{329--342}.
\newblock
\showISBNx{978-1-4503-0663-8}
\urldef\tempurl%
\url{https://doi.org/10.1145/1993498.1993537}
\showDOI{\tempurl}


\bibitem[\protect\citeauthoryear{Mikelsons}{Mikelsons}{1983}]%
        {mikelsons_interactive_1983}
\bibfield{author}{\bibinfo{person}{Martin Mikelsons}.}
  \bibinfo{year}{1983}\natexlab{}.
\newblock \showarticletitle{Interactive program execution in {Lispedit}}. In
  \bibinfo{booktitle}{\emph{Proceedings of the symposium on {High}-level
  debugging}} \emph{(\bibinfo{series}{{SIGSOFT} '83})}.
  \bibinfo{publisher}{Association for Computing Machinery},
  \bibinfo{address}{New York, NY, USA}, \bibinfo{pages}{71--80}.
\newblock
\showISBNx{978-0-89791-111-5}
\urldef\tempurl%
\url{https://doi.org/10.1145/1006147.1006164}
\showDOI{\tempurl}


\bibitem[\protect\citeauthoryear{Miller and Myers}{Miller and Myers}{2001}]%
        {miller_interactive_2001}
\bibfield{author}{\bibinfo{person}{Robert~C. Miller} {and} \bibinfo{person}{B.
  Myers}.} \bibinfo{year}{2001}\natexlab{}.
\newblock \showarticletitle{Interactive {Simultaneous} {Editing} of {Multiple}
  {Text} {Regions}}. In \bibinfo{booktitle}{\emph{{USENIX} {Annual} {Technical}
  {Conference}, {General} {Track}}}.
\newblock


\bibitem[\protect\citeauthoryear{Nakazawa}{Nakazawa}{2022}]%
        {nakazawa_cpojerjs-codemod_2022}
\bibfield{author}{\bibinfo{person}{Christoph Nakazawa}.}
  \bibinfo{year}{2022}\natexlab{}.
\newblock \bibinfo{title}{cpojer/js-codemod}.
\newblock
\newblock
\urldef\tempurl%
\url{https://github.com/cpojer/js-codemod}
\showURL{%
\tempurl}
\newblock
\shownote{original-date: 2015-03-23T04:45:13Z.}


\bibitem[\protect\citeauthoryear{Ni, Sunshine, Le, Gulwani, and Barik}{Ni
  et~al\mbox{.}}{2021}]%
        {ni_recode_2021}
\bibfield{author}{\bibinfo{person}{Wode Ni}, \bibinfo{person}{Joshua Sunshine},
  \bibinfo{person}{Vu Le}, \bibinfo{person}{Sumit Gulwani}, {and}
  \bibinfo{person}{Titus Barik}.} \bibinfo{year}{2021}\natexlab{}.
\newblock \showarticletitle{{reCode} : {A} {Lightweight} {Find}-and-{Replace}
  {Interaction} in the {IDE} for {Transforming} {Code} by {Example}}. In
  \bibinfo{booktitle}{\emph{The 34th {Annual} {ACM} {Symposium} on {User}
  {Interface} {Software} and {Technology}}}. \bibinfo{publisher}{ACM},
  \bibinfo{address}{Virtual Event USA}, \bibinfo{pages}{258--269}.
\newblock
\showISBNx{978-1-4503-8635-7}
\urldef\tempurl%
\url{https://doi.org/10.1145/3472749.3474748}
\showDOI{\tempurl}


\bibitem[\protect\citeauthoryear{Omar, Voysey, Chugh, and Hammer}{Omar
  et~al\mbox{.}}{2019}]%
        {omar_live_2019}
\bibfield{author}{\bibinfo{person}{Cyrus Omar}, \bibinfo{person}{Ian Voysey},
  \bibinfo{person}{Ravi Chugh}, {and} \bibinfo{person}{Matthew~A. Hammer}.}
  \bibinfo{year}{2019}\natexlab{}.
\newblock \showarticletitle{Live functional programming with typed holes}.
\newblock \bibinfo{journal}{\emph{Proceedings of the ACM on Programming
  Languages}} \bibinfo{volume}{3}, \bibinfo{number}{POPL} (\bibinfo{date}{Jan.}
  \bibinfo{year}{2019}), \bibinfo{pages}{14:1--14:32}.
\newblock
\urldef\tempurl%
\url{https://doi.org/10.1145/3290327}
\showDOI{\tempurl}


\bibitem[\protect\citeauthoryear{Pike}{Pike}{1987}]%
        {pike_structural_1987}
\bibfield{author}{\bibinfo{person}{Rob Pike}.} \bibinfo{year}{1987}\natexlab{}.
\newblock \bibinfo{booktitle}{\emph{Structural {Regular} {Expressions}}}.
\newblock \bibinfo{type}{{T}echnical {R}eport}. \bibinfo{institution}{AT\&T
  Bell Laboratories}.
\newblock


\bibitem[\protect\citeauthoryear{Safronov}{Safronov}{2021}]%
        {safronov_awesome_2021}
\bibfield{author}{\bibinfo{person}{Yevgen Safronov}.}
  \bibinfo{year}{2021}\natexlab{}.
\newblock \bibinfo{title}{awesome jscodeshift}.
\newblock
\newblock
\urldef\tempurl%
\url{https://github.com/sejoker/awesome-jscodeshift}
\showURL{%
\tempurl}
\newblock
\shownote{original-date: 2016-03-05T21:07:18Z.}


\bibitem[\protect\citeauthoryear{Sveen}{Sveen}{2021}]%
        {sveen_multiple-cursors_2021}
\bibfield{author}{\bibinfo{person}{Magnar Sveen}.}
  \bibinfo{year}{2021}\natexlab{}.
\newblock \bibinfo{title}{multiple-cursors.el}.
\newblock
\newblock
\urldef\tempurl%
\url{https://github.com/magnars/multiple-cursors.el}
\showURL{%
\tempurl}
\newblock
\shownote{original-date: 2012-01-24T08:45:50Z.}


\bibitem[\protect\citeauthoryear{Teitelbaum and Reps}{Teitelbaum and
  Reps}{1981}]%
        {teitelbaum_cornell_1981}
\bibfield{author}{\bibinfo{person}{Tim Teitelbaum} {and}
  \bibinfo{person}{Thomas Reps}.} \bibinfo{year}{1981}\natexlab{}.
\newblock \showarticletitle{The {Cornell} program synthesizer: a
  syntax-directed programming environment}.
\newblock \bibinfo{journal}{\emph{Commun. ACM}} \bibinfo{volume}{24},
  \bibinfo{number}{9} (\bibinfo{date}{Sept.} \bibinfo{year}{1981}),
  \bibinfo{pages}{563--573}.
\newblock
\showISSN{0001-0782, 1557-7317}
\urldef\tempurl%
\url{https://doi.org/10.1145/358746.358755}
\showDOI{\tempurl}


\end{thebibliography}

\addtolength{\textheight}{3.02cm}

\onecolumn
\clearpage
\pagebreak
\edef\thefigurebeforeappendix{\thefigure}
\appendix
\twocolumn
\raggedbottom
\section{Appendix}
\subsection{Basic Commands in Forest}
\label{sec:single-cursor-commands}
The following commands are used for both single-cursor and multi-cursor editing in Forest. Section \ref{sec:single-cursor} describes the general behavior of Forest, as well as the design decisions concerning some of the following commands.
\\
\\
\keys{\arrowkeyup}\hspace{1.5mm}
Move to parent
\vspace{1mm}\\ 
\keys{\arrowkeyleft \space\slash\space \arrowkeyright}\hspace{1.5mm}
Move to previous\slash{}next leaf node
\vspace{1mm}\\
\keys{\shift + \arrowkeyleft \space\slash\space \arrowkeyright}\hspace{1.5mm}
Extend selection to previous\slash{}next leaf node
\vspace{1mm}\\
\keys{\SPACE}\hspace{1.5mm}
Reduce selection to element just added by extend
\vspace{1mm}\\
\keys{\Alt + \arrowkeyleft \space\slash\space \arrowkeyright}\hspace{1.5mm}
Reduce selection to first\slash{}last element
\vspace{1mm}\\
\keys{(}, \keys{[}, \keys{\{}, \keys{<}\hspace{1.5mm}
Select contents of first list delimited by this matching pair (descendant of current selection)
\vspace{1mm}\\
\keys{\arrowkeydown}\hspace{1.5mm}
Select contents of first list delimited by any matching pair (descendant of current selection)
\vspace{1mm}\\
\keys{)}, \keys{]}, \keys{\}}, \keys{>}\hspace{1.5mm}
Select closest list delimited by this matching pair (ancestor of current selection)
\vspace{1mm}\\
\keys{\shift + \arrowkeyup}\hspace{1.5mm}
Select closest list delimited by any matching pair (ancestor of current selection)
\vspace{1mm}\\
\keys{z}, \keys{\shift + z}\hspace{1.5mm}
Undo or redo selection change
\vspace{1mm}\\
\keys{\ctrl + \shift + \arrowkeyleft \space\slash\space \arrowkeyright}\hspace{1.5mm}
Remove last\slash{}first element from selection
\vspace{1mm}\\
\keys{i}, \keys{a}\hspace{1.5mm}
Insert text before/after cursor
\vspace{1mm}\\
\keys{d}\hspace{1.5mm}
Delete selected nodes
\vspace{1mm}\\
\keys{c}, \keys{p}\hspace{1.5mm}
Copy and paste

\subsection{Multi-Cursor Commands in Forest}
The following commands are used for performing multi-cursor editing. Their behavior is described in Section \ref{sec:multi-cursor}.
\\
\\
\keys{s}\hspace{1.5mm}
Split cursor by creating cursors for each selected list item
\vspace{1mm}\\
\keys{q}\hspace{1.5mm}
Queue selection to later create a cursor with
\vspace{1mm}\\
\keys{\shift + q}\hspace{1.5mm} Create cursors from each queued selection (replaces existing cursor)
\vspace{1mm}\\
\keys{\shift + s} \keys{\arrowkeyleft \space\slash\space \arrowkeyright \space\slash\space\space \arrowkeyup \space\space\slash\space\space \arrowkeydown}\hspace{1.5mm}
Remove all cursors except the first\slash{}last\slash{}outermost\slash{}innermost ones
\vspace{1mm}\\
\keys{m} \keys{letter}\hspace{1.5mm}
Save current selection as mark (named by letter)
\vspace{1mm}\\
\keys{\shift + m} \keys{letter}\hspace{1.5mm}
Jump to selection that was saved as mark (named by letter)
\vspace{1mm}\\
\keys{r}\hspace{1.5mm}
Rename all selected identifiers using JavaScript expression
\vspace{1mm}\\
\keys{\slash}\hspace{1.5mm}
Open structural search
\vspace{1mm}\\
\keys{y} \keys{r \slash\space d \slash\space s}\hspace{1.5mm} Change multi-cursor mode to relaxed\slash{}drop\slash{}strict
\vspace{1mm}\\
\keys{\shift + y} \keys{s \slash\space f \slash\space a}\hspace{1.5mm} Restore state before failure and keep successful\slash{}failed\slash{}all cursors (branching)
\vspace{1mm}\\
\keys{\shift + y} \keys{i}\hspace{1.5mm}
Ignore failure (keep current state and cursors)

\changed{\subsection{Listing of Ignored Repositories}
\label{sec:listing-ignored-repos}
\subsubsection*{\texttt{AMD Transformer}}
The scripts in this repository modify a codebase to use the AMD module system. We were not familiar enough with the AMD module system to work with these scripts.
\subsubsection*{\texttt{coffee-to-es2015-codemod}}
This repository contains a decompiler written in the form of refactoring scripts. The scripts read the output of the CoffeeScript compiler, guess the high-level constructs which were likely used in the original CoffeeScript code, and replace them by their JavaScript equivalent.
\subsubsection*{\texttt{lodash-to-lodash-amd-codemods}}
The scripts in this repository modify the way that functions from a specific library are imported and called, in order for the final program to be efficiently processed by a build system. The repository did not contain sufficient examples to understand the scripts without knowledge of this specific module loading system.
\subsubsection*{\texttt{react-codemod}}
The scripts in this repository perform changes which are helpful when migrating applications to newer versions of the React framework. To ensure that they work robustly in large codebases, most of the scripts handle multiple special cases and contain extensive checks. Handling this many special cases and checks is out of scope for our system.}

\changed{\subsection{Listing of Refactoring Scripts}
\label{sec:listing-scripts}
\subsubsection*{\texttt{5to6-codemod}}
\paragraph{\texttt{amd}}\hangindent=\parindent\mbox{}\newline{}Result: \emph{no}; Issues: A9, D1\newline{}Compile AMD modules to ES6 modules.
\paragraph{\texttt{cjs}}\hangindent=\parindent\mbox{}\newline{}Result: \emph{no}; Issues: C1, D1\newline{}Compile CommonJS modules to ES6 modules (imports only).
\paragraph{\texttt{exports}}\hangindent=\parindent\mbox{}\newline{}Result: \emph{no}; Issues: C1\newline{}Compile CommonJS modules to ES6 modules (exports only).
\paragraph{\texttt{let}}\hangindent=\parindent\mbox{}\newline{}Result: \emph{yes}; Issues: none\newline{}Replace \texttt{var} with \texttt{let}, independent of how the variable is used.
\paragraph{\texttt{named-export-generation}}\hangindent=\parindent\mbox{}\newline{}Result: \emph{no}; Issues: C1\newline{}Generate a named export for every property of a default-exported object literal.
\paragraph{\texttt{no-strict}}\hangindent=\parindent\mbox{}\newline{}Result: \emph{yes}; Issues: none\newline{}Delete a specific statement.
\paragraph{\texttt{simple-arrow}}\hangindent=\parindent\mbox{}\newline{}Result: \emph{maybe}; Issues: A3\newline{}Replace function expressions with a single return by arrow functions with shorthand returns.
\subsubsection*{\texttt{es5-function-to-class-codemod}}
\paragraph{\texttt{func-to-class}}\hangindent=\parindent\mbox{}\newline{}Result: \emph{no}; Issues: B7, D1\newline{}Convert classes declared with functions and prototypes to ES6 classes.
\subsubsection*{\texttt{js-codemod}}
\paragraph{\texttt{arrow-function}}\hangindent=\parindent\mbox{}\newline{}Result: \emph{maybe}; Issues: A1, A2, B2\newline{}Replace normal functions with arrow functions with some exceptions. The functions must be unbound when or bound only with \texttt{this}. The body of unbound functions must not reference \texttt{this}.
\paragraph{\texttt{arrow-function-arguments}}\hangindent=\parindent\mbox{}\newline{}Result: \emph{no}; Issues: A1, B7, C1\newline{}Create an array expression which represents all the arguments to an arrow function. Add a spread to the parameter list if necessary. Replace all usages of the variable \texttt{arguments} by that expression.
\paragraph{\texttt{expect}}\hangindent=\parindent\mbox{}\newline{}Result: \emph{no}; Issues: C2\newline{}Switch every assertion from one assertion library to another. Replace certain function names. Sometimes replace the whole expression and copy over parts.
\paragraph{\texttt{flow-bool-to-boolean}}\hangindent=\parindent\mbox{}\newline{}\emph{This script was ignored}\newline{}Replace references to one type by references to another type. Only modify references that are type position. This script uses non-standard syntax.
\paragraph{\texttt{invalid-requires}}\hangindent=\parindent\mbox{}\newline{}\emph{This script was ignored}\newline{}Same as \texttt{unchain-variables}, but limited to statements that contain \texttt{require}. This script was ignored, because it is too similar to the other script.
\paragraph{\texttt{jest-11-update}}\hangindent=\parindent\mbox{}\newline{}\emph{This script was ignored}\newline{}No example code.
\paragraph{\texttt{jest-arrow}}\hangindent=\parindent\mbox{}\newline{}Result: \emph{maybe}; Issues: A3\newline{}Replace normal functions with arrow functions with some exceptions. Only modify function that are used as arguments to calls to a specific library.
\paragraph{\texttt{jest-remove-describe}}\hangindent=\parindent\mbox{}\newline{}Result: \emph{maybe}; Issues: B4, B9\newline{}Replace function calls by their body with some exceptions. Only modify functions that have a specific name. Only modify functions that are at the top level of the program.
\paragraph{\texttt{jest-remove-disable-automock}}\hangindent=\parindent\mbox{}\newline{}Result: \emph{yes}; Issues: none\newline{}Remove calls to a specific function. Preserve any calls that are chained from the removed call.
\paragraph{\texttt{jest-rm-mock}}\hangindent=\parindent\mbox{}\newline{}\emph{This script was ignored}\newline{}No example code.
\paragraph{\texttt{jest-update}}\hangindent=\parindent\mbox{}\newline{}\emph{This script was ignored}\newline{}No example code.
\paragraph{\texttt{no-reassign-params}}\hangindent=\parindent\mbox{}\newline{}\emph{This script was ignored}\newline{}No example code.
\paragraph{\texttt{no-vars}}\hangindent=\parindent\mbox{}\newline{}Result: \emph{no}; Issues: C1\newline{}Replace \texttt{var} with \texttt{let} or \texttt{const} depending on usage.
\paragraph{\texttt{object-shorthand}}\hangindent=\parindent\mbox{}\newline{}Result: \emph{no}; Issues: B6\newline{}Convert properties where both sides are the same to shorthand syntax. Convert properties whose value is a function to method syntax.
\paragraph{\texttt{outline-require}}\hangindent=\parindent\mbox{}\newline{}\emph{This script was ignored}\newline{}No example code.
\paragraph{\texttt{rm-copyProperties}}\hangindent=\parindent\mbox{}\newline{}Result: \emph{no}; Issues: C1\newline{}Replace calls to a library function with \texttt{Object.assign} or object literals. This script has many checks and exceptions.
\paragraph{\texttt{rm-merge}}\hangindent=\parindent\mbox{}\newline{}Result: \emph{yes}; Issues: A1, A4\newline{}Replace calls to library function with object literals.
\paragraph{\texttt{rm-object-assign}}\hangindent=\parindent\mbox{}\newline{}Result: \emph{maybe}; Issues: A5\newline{}Replace calls to `Object.assign` with object literals. Only modify calls where the first argument is a literal. Do not modify calls where any argument is spread.
\paragraph{\texttt{rm-requires}}\hangindent=\parindent\mbox{}\newline{}Result: \emph{no}; Issues: B1, B5\newline{}Remove calls to specific function if their result is never used. Remove duplicate function calls and update references to their results.
\paragraph{\texttt{template-literals}}\hangindent=\parindent\mbox{}\newline{}Result: \emph{no}; Issues: D1\newline{}Replace additions of strings with template literals.
\paragraph{\texttt{touchable}}\hangindent=\parindent\mbox{}\newline{}Result: \emph{no}; Issues: D1\newline{}Replace JSX element by its children. Only replace elements that have a specific kind of parent. Only replace elements that are the only child of their parent.
\paragraph{\texttt{trailing-commas}}\hangindent=\parindent\mbox{}\newline{}\emph{This script was ignored}\newline{}This script only affects code formatting.
\paragraph{\texttt{unchain-variables}}\hangindent=\parindent\mbox{}\newline{}Result: \emph{yes}; Issues: none\newline{}Flatten variable declarations into multiple statements.
\paragraph{\texttt{underscore-to-lodash-native}}\hangindent=\parindent\mbox{}\newline{}Result: \emph{no}; Issues: C2\newline{}Replace calls to library with calls to other library or built in functions. This script affects calls to many different functions.
\paragraph{\texttt{unquote-properties}}\hangindent=\parindent\mbox{}\newline{}\emph{This script was ignored}\newline{}Remove quotes from property names if they are not necessary. Quoted properties are not supported by our prototype.
\paragraph{\texttt{updated-computed-props}}\hangindent=\parindent\mbox{}\newline{}\emph{This script was ignored}\newline{}No example code.
\paragraph{\texttt{use-strict}}\hangindent=\parindent\mbox{}\newline{}Result: \emph{maybe}; Issues: B3\newline{}Add a specific statement to the start of each file.

\addtolength{\textheight}{-3cm}

\subsubsection*{\texttt{js-transforms}}
\paragraph{\texttt{bind-this-to-bind-expression}}\hangindent=\parindent\mbox{}\newline{}\emph{This script was ignored}\newline{}Replace specific function calls by non-standard syntax.
\paragraph{\texttt{call-expression-bind-...-function-expression}}\hangindent=\parindent\mbox{}\newline{}Result: \emph{maybe}; Issues: A1, A3, A4, A8\newline{}Replace function expressions by arrow functions. Replace arrow function bodies containing single return by shorthand. Only modify functions that are directly used with \texttt{.bind(this)}.
\paragraph{\texttt{function-expression-...-function-expression}}\hangindent=\parindent\mbox{}\newline{}Result: \emph{maybe}; Issues: none\newline{}Replace function expressions by arrow functions. Replace arrow function bodies containing single return by shorthand. Only modify functions whose body does not contain \texttt{this}.
\paragraph{\texttt{props-to-destructuring}}\hangindent=\parindent\mbox{}\newline{}Result: \emph{no}; Issues: B5, B7, C1, C2\newline{}Replace property accesses by destructuring and variable references. Do not add a destructure if the variable already exits. Do not use any reserved words.
\paragraph{\texttt{pure-to-composite-component}}\hangindent=\parindent\mbox{}\newline{}Result: \emph{maybe}; Issues: A1, A3, B2, D1\newline{}Replace an expression and copy over some parts. Only modify expressions that contain specific syntax. Replace references to a specific variable by property access.
\subsubsection*{\texttt{mocha2ava-codemod}}
\paragraph{\texttt{add-pass-test}}\hangindent=\parindent\mbox{}\newline{}Result: \emph{maybe}; Issues: A2, A2, B1\newline{}Modify functions that are used as arguments to a specific function. Applies to arrow functions, normal functions and functions wrapped in a call. Append a statement to the end of the function. Only modify functions if a specific variable is never used inside their body.
\paragraph{\texttt{extractDescribes}}\hangindent=\parindent\mbox{}\newline{}Result: \emph{no}; Issues: A7, C3\newline{}Flatten a set of nested functions and function calls. Concatenate strings used as arguments in each flattened level.
\paragraph{\texttt{insertRequires}}\hangindent=\parindent\mbox{}\newline{}Result: \emph{no}; Issues: A1, D1\newline{}Add an import statement or a variable declaration with a function call. Do not add the statement if it already exists. Use an import statement if there are any other import statements in the file.
\paragraph{\texttt{it2test}}\hangindent=\parindent\mbox{}\newline{}Result: \emph{maybe}; Issues: none\newline{}Add a property access before a function call. Add an argument to the callback (arrow function or function expression). Only modify calls to a specific set of functions. Replace calls to a specific function by calls to a different function.
\paragraph{\texttt{this2context}}\hangindent=\parindent\mbox{}\newline{}Result: \emph{yes}; Issues: none\newline{}Replace the left hand side of a property access by another property access. Only modify property accesses whose left hand side is \texttt{this}.
\subsubsection*{\texttt{preact-codemod}}
\paragraph{\texttt{component}}\hangindent=\parindent\mbox{}\newline{}Result: \emph{maybe}; Issues: A1\newline{}Apply one of two other refactorings to calls to a specific function.
\paragraph{\texttt{component-class}}\hangindent=\parindent\mbox{}\newline{}Result: \emph{maybe}; Issues: D1\newline{}Replace a function call with an object literal with functions by a class with methods.
\paragraph{\texttt{component-sfc}}\hangindent=\parindent\mbox{}\newline{}Result: \emph{yes}; Issues: none\newline{}Replace a function call by a function expression from one of the arguments. Only modify function calls whose argument is an object with certain properties.
\paragraph{\texttt{import-declarations}}\hangindent=\parindent\mbox{}\newline{}Result: \emph{maybe}; Issues: A1, B3\newline{}Replace a specific import by another import. Add an import if a specific function is used.
\paragraph{\texttt{props}}\hangindent=\parindent\mbox{}\newline{}Result: \emph{yes}; Issues: A3, D1\newline{}Replace property accesses by variable references. Add an argument to the containing function. Only modify the containing function if it has no arguments.
\paragraph{\texttt{removePropTypes}}\hangindent=\parindent\mbox{}\newline{}Result: \emph{yes}; Issues: none\newline{}Remove a specific import. Remove assignments to a specific property.
\paragraph{\texttt{state}}\hangindent=\parindent\mbox{}\newline{}Result: \emph{yes}; Issues: none\newline{}Replace property accesses by variable references. Add an argument to the containing function. Only modify the containing function if it has no arguments. Add another argument to the containing function if it has one argument.
\subsubsection*{\texttt{rackt-codemod}}
\paragraph{\texttt{deprecate-createPath-createHref-query}}\hangindent=\parindent\mbox{}\newline{}Result: \emph{yes}; Issues: none\newline{}Replace arguments to a specific function by named arguments in an object literal. Replace strings by expressions created by parsing the strings.
\paragraph{\texttt{deprecate-isActive-query}}\hangindent=\parindent\mbox{}\newline{}Result: \emph{no}; Issues: A2, A5, B8\newline{}Replace arguments to a specific function by named arguments in an object literal. Different handling depending on number of arguments.
\paragraph{\texttt{deprecate-pushState-replaceState}}\hangindent=\parindent\mbox{}\newline{}\emph{This script was ignored}\newline{}Replace arguments to specific function by named arguments in an object literal. Replace strings by expressions created by parsing the strings.
\paragraph{\texttt{react-router/deprecate-Link-location-props}}\hangindent=\parindent\mbox{}\newline{}Result: \emph{no}; Issues: D1\newline{}Modify a JSX prop if another prop from a specific set is present. Copy over expressions and delete old props.
\paragraph{\texttt{react-router/deprecate-context-history}}\hangindent=\parindent\mbox{}\newline{}Result: \emph{maybe}; Issues: A2\newline{}Multiple different structural find-replaces.
\subsubsection*{\texttt{rm-debugger}}
\paragraph{\texttt{rm-debugger}}\hangindent=\parindent\mbox{}\newline{}\emph{This script was ignored}\newline{}Remove a specific statement. This script is unrealistically simple.
\subsubsection*{\texttt{undecorate-codemod}}
\paragraph{\texttt{undecorate}}\hangindent=\parindent\mbox{}\newline{}Result: \emph{no}; Issues: D1\newline{}Remove decorators from class declarations and wrap them in a corresponding function call. Create a temporary variable and export statement if the class was not default-exported.
\subsubsection*{\texttt{vue-codemods}}
\paragraph{\texttt{extract\_non\_instance\_methods}}\hangindent=\parindent\mbox{}\newline{}Result: \emph{maybe}; Issues: B1, B1, D1\newline{}Replace methods in an object literal by functions at the top level of the file.
\paragraph{\texttt{sort\_keys}}\hangindent=\parindent\mbox{}\newline{}Result: \emph{no}; Issues: B11\newline{}Sort properties within an object literal according to multiple rules. Sort certain properties according to a lookup table.
\paragraph{\texttt{uppercase\_constants}}\hangindent=\parindent\mbox{}\newline{}Result: \emph{no}; Issues: A2, B1, B10, B4, B6\newline{}Rename variable names to upper case in \texttt{const} declarations. Rename all references to the renamed variable. Replace string constants by identifiers and create variable declarations if the same constant appears multiple times.
\subsubsection*{\texttt{webpack-babel-codemod}}
\paragraph{\texttt{dynamic-require-import}}\hangindent=\parindent\mbox{}\newline{}Result: \emph{maybe}; Issues: A6, D1\newline{}Replace properties whose value is a function call by shorthand properties. Hoist function calls to the top of the file.}

\begin{table*}[ht]
\caption{Issues encountered while performing the edits of real-world refactoring scripts (Section \ref{sec:evaluation-refactoring-scripts}). This is a full version of Table \ref{table:refactoring-script-issues} with no issues omitted. The numbers in each row indicate the number of scripts where we encountered the given issue. The \emph{Total} column counts scripts regardless of how they were classified. The other columns only count scripts that had the corresponding classification. For example, we encountered the issue ``Have to recreate cursors multiple times'' with 8 scripts, 2 of which were classified \emph{No}. Note that each script may have multiple issues. Scripts classified \emph{Yes} often had no issues. The issues (rows in \emph{italics}) are grouped into categories (rows in \textbf{bold}).}
\label{table:refactoring-script-issues-full}
\begin{tabular}{@{}cccccl@{}} \toprule
\multicolumn{4}{c}{Affected Scripts} \\\cmidrule(r){2-5}
\changed{ID} & Total & No & Maybe & Yes & \textbf{Category} or \emph{Specific Issue} \\\midrule
& \textbf{27} & \textbf{7} & \textbf{17} & \textbf{3} & \textbf{Missing features that require conceptual changes} \\
\changed{A1} & \emph{8} & \emph{2} & \emph{5} & \emph{1} & \emph{Have to recreate cursors multiple times} \\
\changed{A2} & \emph{6} & \emph{2} & \emph{4} & \emph{0} & \emph{Cannot handle separately found locations together} \\
\changed{A3} & \emph{5} & \emph{0} & \emph{4} & \emph{1} & \emph{Nested copy-paste would be an issue} \\
\changed{A4} & \emph{2} & \emph{0} & \emph{1} & \emph{1} & \emph{Manual parenthesizing required} \\
\changed{A5} & \emph{2} & \emph{1} & \emph{1} & \emph{0} & \emph{None-one-many issue} \\
\changed{A6} & \emph{1} & \emph{0} & \emph{1} & \emph{0} & \emph{Adding after first import \emph{or} as first statement if no imports doesn't work} \\
\changed{A7} & \emph{1} & \emph{1} & \emph{0} & \emph{0} & \emph{Can do the flatten sometimes, but not in the general case} \\
\changed{A8} & \emph{1} & \emph{0} & \emph{1} & \emph{0} & \emph{Matching is different because of search in flattened AST} \\
\changed{A9} & \emph{1} & \emph{1} & \emph{0} & \emph{0} & \emph{No support for zipping lists of cursors} \\
\hline
& \textbf{21} & \textbf{12} & \textbf{9} & \textbf{0} & \textbf{Missing features that do not require conceptual changes} \\
\changed{B1} & \emph{5} & \emph{2} & \emph{3} & \emph{0} & \emph{No strict ``find usages of variable''} \\
\changed{B2} & \emph{2} & \emph{0} & \emph{2} & \emph{0} & \emph{Bug in paste} \\
\changed{B3} & \emph{2} & \emph{0} & \emph{2} & \emph{0} & \emph{Cannot edit multiple files} \\
\changed{B4} & \emph{2} & \emph{1} & \emph{1} & \emph{0} & \emph{Cannot filter top-level statements} \\
\changed{B5} & \emph{2} & \emph{2} & \emph{0} & \emph{0} & \emph{Cannot remove duplicate items} \\
\changed{B6} & \emph{2} & \emph{2} & \emph{0} & \emph{0} & \emph{Cannot search for dynamic query} \\
\changed{B7} & \emph{2} & \emph{2} & \emph{0} & \emph{0} & \emph{No strict ``find declarations of variable''} \\
\changed{B8} & \emph{1} & \emph{1} & \emph{0} & \emph{0} & \emph{Cannot filter for exactly one item in list} \\
\changed{B9} & \emph{1} & \emph{0} & \emph{1} & \emph{0} & \emph{Cannot filter for exactly one search match} \\
\changed{B10} & \emph{1} & \emph{1} & \emph{0} & \emph{0} & \emph{Cannot search up} \\
\changed{B11} & \emph{1} & \emph{1} & \emph{0} & \emph{0} & \emph{No sort feature} \\
\hline
& \textbf{11} & \textbf{11} & \textbf{0} & \textbf{0} & \textbf{Edit is too complicated for multi-cursor structural editing} \\
\changed{C1} & \emph{7} & \emph{7} & \emph{0} & \emph{0} & \emph{Too complicated} \\
\changed{C2} & \emph{3} & \emph{3} & \emph{0} & \emph{0} & \emph{Lookup tables are possible but impractical} \\
\changed{C3} & \emph{1} & \emph{1} & \emph{0} & \emph{0} & \emph{Cannot create AST by parsing arbitrary string using JS} \\
\hline
& \textbf{11} & \textbf{6} & \textbf{4} & \textbf{1} & \textbf{Unsupported syntax} \\
\changed{D1} & \emph{11} & \emph{6} & \emph{4} & \emph{1} & \emph{Unsupported syntax} \\
\bottomrule
\vspace{0.1cm}
\end{tabular}
\end{table*}

\onecolumn
\clearpage
\pagebreak
\twocolumn

\end{document}